\documentclass[11pt]{article}
 \usepackage{titlesec}
\usepackage{hyperref}

\titleclass{\subsubsubsection}{straight}[\subsection]

\usepackage[utf8]{inputenc}
\usepackage{amsmath,amsfonts,amssymb,stackengine,graphicx}
\title{appendix H2}
\author{sagnotti }
\date{June 2020}

\usepackage[utf8]{inputenc}
\newif\iffigs\figstrue

\usepackage[title]{appendix}
\usepackage{epsfig,latexsym}
\usepackage{hyperref}
\usepackage{amsmath}
\usepackage{verbatim}
\usepackage{color}
\usepackage{mathrsfs}  
\usepackage{slashed}
\usepackage{amssymb}

\DeclareMathAlphabet{\mathpzc}{OT1}{pzc}{m}{it}

 \csname
@addtoreset\endcsname{equation}{section}
  
\def\gz0{\gamma^{0}}

\def\sign{\rm sign}

\def\beq{\begin{equation}}
\def\eeq{\end{equation}}
\def\bea{\begin{eqnarray}}
\def\eea{\end{eqnarray}}
\def\ba{\begin{array}}
\def\ea{\end{array}}
\def\bec{\begin{center}}
\def\ec{\end{center}}
\def\ba{\begin{align}}
\def\ena{\end{align}}

\def\12{\frac{1}{2}}

\def\pr{\partial}

\newcounter{subsubsubsection}[subsubsection]
\renewcommand\thesubsubsubsection{\thesubsubsection.\arabic{subsubsubsection}}
\titleformat{\subsubsubsection}
  {\normalfont\normalsize\bfseries}{\thesubsubsubsection}{1em}{}
\titlespacing*{\subsubsubsection}
{0pt}{3.25ex plus 1ex minus .2ex}{1.5ex plus .2ex}

\makeatletter
\renewcommand\paragraph{\@startsection{paragraph}{5}{\z@}%
  {3.25ex \@plus1ex \@minus.2ex}%
  {-1em}%
  {\normalfont\normalsize\bfseries}}
\renewcommand\subparagraph{\@startsection{subparagraph}{6}{\parindent}%
  {3.25ex \@plus1ex \@minus .2ex}%
  {-1em}%
  {\normalfont\normalsize\bfseries}}
\def\toclevel@subsubsubsection{4}
\def\toclevel@paragraph{5}
\def\toclevel@paragraph{6}
\def\l@subsubsubsection{\@dottedtocline{4}{7em}{4em}}
\def\l@paragraph{\@dottedtocline{5}{10em}{5em}}
\def\l@subparagraph{\@dottedtocline{6}{14em}{6em}}
\makeatother

\begin{document}

\begin{flushright}
{\today}
\end{flushright}

\vspace{10pt}

\begin{center}

%%%%%%%%%%%%%%%%%%%%%%%%%%%%%%%%%%%%%%%%%%%%%%%%%%%%%%%%%%%%%%%%%%%%

{\Large\sc On Warped String Vacuum Profiles and Cosmologies, I}\\
{\large\sc Supersymmetric Strings}

%%%%%%%%%%%%%%%%%%%%%%%%%%%%%%%%%%%%%%%%%%%%%%%%%%%%%%%%%%%%%%%%%%%%

\vspace{25pt}
{\sc J.~Mourad${}^{\; a}$  \ and \ A.~Sagnotti${}^{\; b}$\\[15pt]

${}^a$\sl\small APC, UMR 7164-CNRS\\
Universit\'e de Paris  \\
10 rue Alice Domon et L\'eonie Duquet \\75205 Paris Cedex 13 \ FRANCE
\\ e-mail: {\small \it
mourad@apc.univ-paris7.fr}\vspace{10
pt}

{${}^b$\sl\small
Scuola Normale Superiore and INFN\\
Piazza dei Cavalieri, 7\\ 56126 Pisa \ ITALY \\
e-mail: {\small \it sagnotti@sns.it}}\vspace{10pt}
}

%%%%%%%%%%%%%%%%%%%%%%%%%%%%%%%%%%%%%%%%%%%%%%%
\vspace{40pt} {\sc\large Abstract}\end{center}
%%%%%%%%%%%%%%%%%%%%%%%%%%%%%%%%%%%%%%%%%%%%%%%
\noindent
We investigate in detail solutions of supergravity that involve warped products of flat geometries of the type $M_{p+1} \times R \ \times T_{D-p-2}$ depending on a single coordinate. In the absence of fluxes, the solutions include flat space and Kasner--like vacua that break all supersymmetries. In the presence of a symmetric flux, there are three families of solutions that are characterized by a pair of boundaries and have a singularity at one of them, the origin. The first family comprises supersymmetric vacua, which capture a universal limiting behavior at the origin. The first and second families also contain non--supersymmetric solutions whose behavior at the other boundary, which can lie at a finite or infinite distance, is captured by the no--flux solutions. The solutions of the third family have a second boundary at a finite distance where they approach again the supersymmetric backgrounds.
These vacua exhibit a variety of interesting scenarios, which include compactifications on finite intervals and $p+1$--dimensional effective theories where the string coupling has an upper bound. We also build corresponding cosmologies, and in some of them the string coupling can be finite throughout the evolution.

\setcounter{page}{1}

\pagebreak

\newpage
\tableofcontents
\newpage
\baselineskip=20pt
%%%%%%%%%%%%%%%%%%%%%%%%%%%%%%%%%%
\section{\sc  Introduction and Summary}\label{sec:intro}
%%%%%%%%%%%%%%%%%%%%%%%%%%%%%%%%%%%%%

\vskip 12pt

Supersymmetry is a spacetime symmetry, whose breaking can be realized in manifold ways~\cite{SUSY}, above and beyond the homogeneous setting at work for the gauge symmetries of the Standard Model of Particle Physics. A number of scenarios where a partial breaking of supersymmetry is induced by compactifications on suitable internal manifolds, where extended objects ($p$--branes and orientifolds) are possibly present, have been widely explored in connection with String Theory~\cite{strings}. Still, one can fairly state that our current understanding of String Theory is confined, to a large extent, to vacua where some residual amount of supersymmetry is preserved. These cases can be investigated by powerful tools, which rest on solving first--order Killing spinor equations (for a review, see~\cite{grana}).

In this paper we explore the option of breaking supersymmetry spontaneously and completely via inhomogeneous vacuum solutions of the ten--dimensional supersymmetric strings. We rely on the complete equations of Supergravity~\cite{SUGRA}, prescribing some symmetry properties. 
In contrast with the well--explored Scherk--Schwarz compactifications~\cite{scherkschwarz}, which rest on modified boundary conditions, here the bulk plays a central role. Drawing some motivation from branes, we consider inhomogeneous geometries depending on a single spatial coordinate. The class of vacua that we address is also relevant to eleven--dimensional supergravity~\cite{cjs}, involves warped products of two symmetric spaces, and is described by metric tensors of the form
\beq
ds^{\,2}\ = \ e^{2A(r)}\, \gamma_{\mu\nu}(x)\, dx^\mu\,dx^\nu \ + \ e^{2B(r)}\,dr^2\ + \ e^{2C(r)}\, \gamma_{mn}(\xi)\, d \xi^m\,d \xi^n \ . \label{metric_sym_intro}
\eeq
Here $\gamma_{\mu\nu}(x)$ and $\gamma_{mn}(\xi)$ describe, in general, maximally symmetric spaces of dimensions $p+1$ and $D-p-2$, with curvatures $k$ and $k'$, and the standard BPS branes in ten dimensions are encompassed by eq.~\eqref{metric_sym_intro} with $k=0$ and $k'=1$, so that $\gamma_{\mu\nu}(x)$ is a Minkowski metric while the $\xi$-coordinates describe a sphere. This motivated us to focus on vacua of this type with $k=k'=0$, which can describe compactifications on products of intervals and internal tori resulting in lower--dimensional Minkowski spaces. This type of setting is relatively simple, and yet it suffices to provide different scenarios where supersymmetry breaking is induced in static vacua, starting from supersymmetric strings, also in the presence of internal fluxes compatible with the symmetries of eq.~\eqref{metric_sym_intro}. Backgrounds of this type can describe compactifications to flat space where gravity and gauge interactions are effectively $p+1$--dimensional, as in~\cite{rs}. In addition, most of the static vacua that we shall construct have interesting cosmological counterparts, which can be reached by analytic continuation.

The plan of this paper is as follows. In Section~\ref{sec:symmetries} we set up our notation, describe the relevant portions of the string effective actions and their main features, together with the salient properties of the class of metrics of interest. We also identify the symmetric field profiles that are allowed in them. In Section~\ref{sec:reduced_action} we build a reduced action principle, and we identify the convenient ``harmonic gauge''
\beq{}{}{}{}
B \ = \ \left(p+1\right) A \ + \ \left(D-p-2\right) C \ , 
\eeq
which greatly simplifies the resulting equations. In Section~\ref{sec:killing_spinors} we determine the supersymmetric vacua of the form~\eqref{metric_sym_intro}, solving the corresponding Killing spinor equations. All these vacua, aside from flat space, require the presence of symmetric fluxes. In Section~\ref{sec:susybnoT} we determine the simplest class of solutions of the form~\eqref{metric_sym_intro} in the absence of internal fluxes. Aside from flat space, they are Kasner--like backgrounds depending on two independent parameters spanning an elliptical cylinder and break supersymmetry completely. We also consider the corresponding cosmologies,  which can be obtained by analytic continuations. In Section~\ref{sec:T0hn0} we determine the general backgrounds of the form~\eqref{metric_sym_intro} in the presence of a symmetric internal form flux. There are three families of solutions, which have a pair of singularities and are distinguished by the energy of the Newtonian model reviewed in Appendix~\ref{app:deq}. The zero--energy family includes the supersymmetric solutions of Section~\ref{sec:killing_spinors}, which capture the limiting behavior at the origin in all cases. The remaining zero--energy solutions and the positive--energy ones approach at the other boundary, which can lie at a finite or infinite distance, the zero--flux Kasner--like backgrounds. Finally, the negative--energy family has a second boundary at a finite distance, where it approaches again the supersymmetric behavior. 
Many new options emerge, including compactifications to $p+1$ dimensions where the string coupling is everywhere bounded. By an analytic continuation, we also build corresponding cosmologies in Section~\ref{sec:cosmoflux}. Our conclusions can be found in Section~\ref{sec:conclusions}. Finally, in Appendix~\ref{app:technical} we collect, for the reader's convenience, some technical details that are used extensively in the paper. 

\vskip 24pt

\section{\sc Effective Action and Symmetric Profiles}\label{sec:symmetries}

In this section we describe our basic setup and the notation that we shall use for the solutions of interest. 
We use a ``mostly--plus'' signature, defining the Riemann curvature tensor via~\footnote{These conventions are as in~\cite{ms_lett_20}.}
\beq
[ \nabla_M, \nabla_N] V_Q \ = \ {R_{MNQ}}^{P}\, V_P \ ,
\eeq
so that
\beq
{R_{MNQ}}^{P} \ = \ \partial_N\,{\Gamma^P}_{MQ} \ - \ \partial_M\,{\Gamma^P}_{NQ}  \ + \ {\Gamma^P}_{NR}\, {\Gamma^R}_{MQ} \ - \ {\Gamma^P}_{MR}\, {\Gamma^R}_{NQ} \ .
\eeq
We also define the Ricci tensor as
\beq
R_{MQ} \ = \ {R_{MNQ}}^{N} \ .
\eeq
In general, late capital Latin indices refer to the collection of curved labels (space--time or internal), and among these we distinguish space--time (small Greek) and internal (late Latin) curved labels. Moreover, all our $\Gamma$--matrices will be curved, while all our $\gamma$--matrices will be flat, and we reserve early Latin letters to flat labels.  

\subsection{\sc Effective Action and Equations of Motion}

In the string frame, the bosonic portions of the low--energy effective field theories that we shall consider, here and in~\cite{ms_vacuum_2}, include the following contributions:
 \bea
 {\cal S} &=&\frac{1}{2\,(\alpha')^\frac{D-2}{2}}\ \int d^{D}x\,\sqrt{-\,G} \Big\{ e^{-2\phi}\Big[\, R\, + \,4(\partial\phi)^2 \Big] \ - \ \frac{\tau_{D-1}}{\alpha'} \, e^{\,\gamma_S\,\phi}\ - \ \frac{1}{2\,(p+2)!}\ e^{-2\,\beta_S\,\phi}\, {\cal H}_{p+2}^2
 \nonumber \\ &-& \frac{\alpha'}{4} \ e^{-2\,\alpha_S\,\phi}\, {\rm tr} \left( {\cal F}_{\mu\nu}\,{\cal F}^{\mu\nu} \right) \Big\} \ . \label{eqs1}
 \eea
This prototype action thus involves, in general, three types of fields aside from gravity: the dilaton $\phi$, a gauge field ${\cal A}$ of field strength
\beq
{\cal F} \ = \ d {\cal A} \ - \ i\, {\cal A}^2 \ ,
\eeq
and a $(p+1)$--form gauge potential ${\cal B}_{p+1}$ of field strength ${\cal H}_{p+2}= d\,{\cal B}_{p+1}$. The values of $p$ and of the two constants $\alpha_S$ and $\beta_S$ for supersymmetric strings are collected in Table~\ref{table:tab_1}. Here the ``tadpole term'' proportional to $\tau_{D-1}$ can describe a non--critical potential if $\gamma_S=-2$ and $D \neq 10$, with $\tau_{D-1} \sim D - 10$, an overall brane--orientifold tension if $\gamma_S=-1$ and $D=10$, or a contribution emerging from genus--one amplitudes if $\gamma_S=0$ and $D=10$. In this paper, our focus will be on supersymmetric critical superstrings and on the eleven--dimensional supergravity, for which $\tau_{D-1}=0$. In~\cite{ms_vacuum_2} we shall study the effects induced by tadpole potentials involving a single exponential, which will depend on the choice of $\gamma$.  Here we shall set up the general formalism, also for generic values of $D$, insofar as possible, before concentrating on the solutions for $\tau_{D-1}=0$ in $D=10,11$.

 \begin{table}
 \begin{center}
\begin{tabular}{ ||c||c|c|c|| } 
 \hline\hline
  Model & $p$ & $\alpha_S$ & $\beta_S$ \\ [0.5ex] 
  \hline\hline
 $IIA$ & $I,V$;$(0,2,4,6,8)$ & -- & $1,\,-1$;\,$(0,0,0,0,0)$ \\ [0.5ex] 
  \hline
 $IIB$ & $I,V$;$(-\,1,1,3,5,7)$ & -- & $1,\,-1$;\,$(0,0,0,0,0)$ \\ [0.5ex] 
  \hline
 $SO(32)$ open & $(1,5)$ & $\frac{1}{2}$ & $(0,0)$ \\ [0.5ex] 
  \hline
 $SO(32)$, $E_8 \times E_8$ heter. & $I,V$ & 1 & $1,\,-1$ \\ [0.5ex] 
  \hline
 \hline\hline
\end{tabular}
 \end{center}
 \vskip 12pt 
 \caption{\small String--frame parameters for the supersymmetric ten--dimensional string models. Roman numerals refer to NS-NS branes, entries within parentheses refer to RR ones, and dashes signal couplings that are not present in the low--energy effective theory.}\vskip 12pt
 \label{table:tab_1}
 \end{table}

In the Einstein frame, with the corresponding metric $g$ related to $G$ according to
\beq
G_{MN} \ = \ e^\frac{4\phi}{D-2} \ g_{MN} \ , \label{string_vs_einstein}
\eeq
the action of eq.~\eqref{eqs1} becomes
\bea
{\cal S} &=& \frac{1}{2(\alpha')^\frac{D-2}{2}}\int d^{D}x\sqrt{-g}\left[R\ - \ \frac{4}{D-2}\ (\partial\phi)^2\ - \ \frac{\tau_{D-1}}{\alpha'} \, e^{\,\gamma\,\phi}\ - \ \frac{e^{-2\,\beta_p\,\phi}}{2\,(p+2)!}\, {\cal H}_{p+2}^2 \right. \nonumber \\
&-& \left. \frac{\alpha'\, e^{-2\,\alpha\,\phi}}{4} \, {\rm tr} \left( {\cal F}_{MN}\,{\cal F}^{MN} \right) \
\right] \ , \label{eqs4}
\eea
with
\beq
\alpha \,=\, \alpha_S \ - \ \frac{D-4}{D-2} \ , \quad
\beta_p \,=\, \beta_S \ - \ \frac{D-2(p+2)}{D-2} \ , \quad
\gamma \,=\, \gamma_S \ + \ \frac{2 D}{D-2} \ .
 \label{alphaE}
\eeq
 \begin{table}
 \begin{center}
\begin{tabular}{ ||c||c|c|c|| } 
 \hline\hline
  Model & $p$ & $\alpha$ & $\beta_p$ \\ [0.5ex] 
  \hline\hline
 $IIA$ & $I,V$;(0,2,4,6,8) & -- & $\frac{1}{2},-\,\frac{1}{2}$;\,$\left(-\,\frac{3}{4},-\,\frac{1}{4},\frac{1}{4},\frac{3}{4},\frac{5}{4}\right)$ \\ [0.6ex] 
  \hline
 $IIB$ & $I,V$;(-1,1,3,5,7) & -- & $\frac{1}{2},-\,\frac{1}{2}$;\,$\left(-\,1,-\,\frac{1}{2},0,\frac{1}{2},1\right)$ \\ [0.6ex] 
  \hline
 $SO(32)$ open & (1,5) & $\frac{1}{2}$ & $\left( -\,\frac{1}{2},\frac{1}{2}\right)$  \\ [0.5ex] 
  \hline
 $SO(32)$, $E_8 \times E_8$ heter.  & $I,V$ & 1 & $\frac{1}{2}$,-\,$\frac{1}{2}$  \\ [0.5ex] 
 \hline\hline
\end{tabular}
 \end{center}
 \vskip 12pt 
 \caption{\small Einstein--frame parameters for the supersymmetric ten--dimensional string models. Roman numerals refer to NS-NS branes, entries within parentheses refer to RR ones, and dashes signal couplings that are not present in the low--energy effective theory.}\vskip 12pt
\label{table:tab_2}
 \end{table}
The values of these quantities for the ten--dimensional superstrings are collected in Table~\ref{table:tab_2}, and the corresponding field equations, here written for a generic potential $V(\phi)$, read
\bea
R_{MN} \,-\, \frac{1}{2}\ g_{MN}\, R &=& \!\!\frac{4}{D-2}\  \pr_M\phi\, \pr_N\phi\, + \, \frac{e^{\,-\,2\,\beta_p\,\phi} }{2(p+1)!}\,  \left({\cal H}_{p+2}^2\right)_{M N} + \frac{\alpha'\,e^{-2\alpha\phi}}{2}\,  {\rm tr} \left( {\cal F}_{M P}\,{{\cal F}_N}^P \right) \nonumber\\
&-& \!\!\frac{1}{2}\,g_{MN}\Big[\frac{4\,(\pr\phi)^2}{D-2}+ \frac{e^{\,-\,2\,\beta_p\,\phi}}{2(p+2)!}\,{\cal H}_{p+2}^2+\frac{\alpha'\,e^{-2\alpha\phi}}{4}\,{\rm tr} \left( {\cal F}_{MN}\,{\cal F}^{MN} \right) + V(\phi)\Big] \ ,  \nonumber
\\
\frac{8}{D-2} \ \Box\phi &=& \,-\,\frac{\beta_p\, e^{\,-\,2\,\beta_p\,\phi}}{(p+2)!} \ {\cal H}_{p+2}^2 \, - \, \frac{\alpha'(D-2)\alpha\,e^{-2\,\alpha\,\phi}}{16} \, {\rm tr} \left( {\cal F}_{MN}\,{\cal F}^{MN} \right) + V^\prime(\phi)  \ , \nonumber 
\\
d\Big(e^{\,-\,2\,\beta_p\,\phi}\ {}^{*}\,{\cal H}_{p+2}\Big) &=& 0 \ , \qquad 
D\Big(e^{\,-\,2\,\alpha\,\phi}\ {}^{*}\,{\cal F}\Big) \ = \  0 \ . \label{eqsbeta}
\eea
We shall often use the alternative form of the Einstein equations
\bea
R_{MN}  &=& \frac{4}{D-2}\  \pr_M\phi\, \pr_N\phi\ + \ \frac{1}{2(p+1)!}\ e^{\,-\,2\,\beta_p\,\phi} \ \left({\cal H}_{p+2}^2\right)_{M N}\,+\,\frac{\alpha'\,e^{-2\alpha\phi}}{2}\,  {\rm tr} \left( {\cal F}_{M P}\,{{\cal F}_N}^P \right) \nonumber\\
&+& g_{MN}\left[\,- \, \frac{(p+1)\, e^{\,-\,2\,\beta_p\,\phi}}{2(D-2)(p+2)!}\, {\cal H}_{p+2}^2\,-\,\frac{\alpha'\,e^{-2\,\alpha\,\phi}}{4(D-2)}\,{\rm tr} \left( {\cal F}_{MN}\,{\cal F}^{MN} \right) \, +\, \frac{V(\phi)}{D-2}\right] \ , \label{eqsnotlagbeta}
\eea
and when taking into account the tadpole potential in~\cite{ms_vacuum_2}, we shall refer to
\beq
V(\phi) \ = \ \frac{\tau_{D-1}}{\alpha'} \ e^{\gamma\,\phi} \ . \label{tadpole_potential}
\eeq
Note also that in the type IIB string there is a five--form field strength, which satisfies the first--order self--duality equation
 \beq{}
{\cal H}_5 \ = \ {}^{*}\,{\cal H}_{5} \ ,
\eeq
and taking it into account will require some slight amendments of the formalism that we shall return to in due time. 

We can now set up our systematic search for symmetric profiles within the class of effective Lagrangians in eq.~\eqref{eqs4}. We begin with a detailed discussion of their features for the various types of fields.

\subsection{\sc Isometry Groups and Metric Profiles}
The familiar supersymmetric $p$--branes live in asymptotically flat $D$--dimensional spacetimes, and are special solutions of the low--energy Supergravity with isometry groups ${ISO(1,p)}\times {SO(D-p-1)}$. These are special cases of more general $D$--dimensional manifolds, with isometry groups ${G_{k}(p+1)}\times { H_{k'}(D-p-2)}$. Here ${G_{k}(p+1)}$ and ${H_{k'}(D-p-2)}$ denote the isometry groups of the maximally symmetric Lorentzian or Euclidean manifolds with dimensions equal to $p+1$ and $D-p-2$, and $k$ and $k'$, equal to $\pm 1$ or $0$, determine the corresponding constant curvatures. As is well known, the space--time isometry groups $G_k$ for the three cases $k=1,0,-1$ are $SO(1,p+1)$, $ISO(1,p)$ and $SO(2,p)$. In a similar fashion, the internal isometry groups $H_{k'}$ for $k'=1,0,-1$ are $SO(D-p-1)$, $ISO(D-p-2)$ and $SO(1,D-p-2)$. In some special cases, as in~\cite{nonsusyads}, the radial coordinate combines with one set or the other, the corresponding symmetry enhances, and different values of $k$ or $k'$ merely reflect different choices of coordinate sets. The ordinary supersymmetric branes have $k=0$ and $k'=1$. 

All the preceding options are available insofar as $p<D-3$, and lead one to distinguish $p+1$ space--time coordinates $x^\mu$, $(\mu=0,\ldots,p)$, $D-p-2$ internal coordinates $\xi^m$, $(m=1,\ldots,D-p-2)$, and a radial coordinate $r$, invariant under the isometry groups, on which the profiles we are after will depend.  If $p=D-3$ there is a single $\xi$--coordinate and only the choice $k'=0$ is possible, while if $p=D-2$ there are no $\xi$--coordinates altogether. We shall work at times with general values of $D$ but, as we have stated, we have in mind primarily critical superstrings, for which $D=10$. The three choices $k=\pm 1,0$ select de Sitter, anti de Sitter and Minkowski space--time manifolds, while $k'=0$ selects a Euclidean internal space (or, more generally, a product of tori, where only continuous internal translations are left), $k'=1$ selects a sphere, while $k'=-1$ selects a hyperbolic space.

The class of $D$--dimensional metrics compatible with the above symmetries takes the form
\beq
ds^{\,2}\ = \ e^{2A(r)}\, \gamma_{\mu\nu}(x)\, dx^\mu\,dx^\nu \ + \ e^{2B(r)}\,dr^2\ + \ e^{2C(r)}\, \gamma_{mn}(\xi)\, d \xi^m\,d \xi^n \ , \label{metric_sym}
\eeq
where $\gamma_{\mu\nu}$ is a $(p+1)$--dimensional metric of constant curvature $k$ and $\gamma_{mn}(\xi)$ is a $(D-p-2)$--dimensional metric of constant curvature $k'$. The complete metric thus involves three dynamical functions of a single variable $r$. The preceding discussion motivates our choice, which was anticipated in eq.~\eqref{metric_sym_intro} of the Introduction.

In this paper we shall begin with the general setup for the field equations to then concentrate on solutions with $k=k'=0$. This choice describes vacua, rather than branes, which include compactifications on internal tori. The redundancy resulting from the introduction of an independent function $B(r)$ has the virtue, as in~\cite{dm_vacuum}, of allowing a wider class of exact solutions. 

Here we shall find it convenient to work in the ``harmonic gauge'', whereby
\beq
B \ = \ (p+1)A \ + \ (D-p-2) C \ , \label{harm_gauge}
\eeq
which will simplify the resulting equations.
Moreover, we shall also explore counterparts of these solutions that are obtained via an analytic continuation of $r$ and $x^0$ to imaginary values. These build anisotropic cosmologies and generalize previous results~\cite{dm_vacuum,russo,dks}~\footnote{In this case the internal symmetry groups remain as above, while $G_k(p+1)$ become $H_k(p+1)$, the isometry groups of the spatial slices of these cosmologies.}. We shall address elsewhere~\cite{ms_vacuum_2} the effects of the tadpole potential~\eqref{tadpole_potential} on vacuum profiles.

\subsection{\sc Symmetric Tensor Profiles} \label{sec:tensor_profiles}

In this section we characterize the symmetric tensor profiles of interest, largely with reference to the case $k=k'=0$, although most expressions have a more general applicability.
Here we are interested in field configurations compatible with the isometries of the class of metrics in eq.~\eqref{metric_sym}. Therefore, we can begin by considering the closed and invariant volume forms~\footnote{In our conventions $\epsilon_{01\ldots}=+1$ and $\epsilon^{01\ldots}=-1$.}
\beq
{\epsilon_{(p+1)}} \ = \ \sqrt{-\gamma(x)} \ dx^0 \wedge ..\wedge dx^{p} \ , \qquad {\widetilde{\epsilon}_{(D-p-2)}}\ = \ \sqrt{\gamma(\xi)} \ d\xi^1 \wedge ..\wedge d\xi^{D-p-2} \ ,
\eeq
since combining them with $dr$ one can the build profiles of closed $r$--dependent form fields
\beq
b'_{p+1}(r)\, {\epsilon_{(p+1)}} \, dr \ , \qquad \tilde{b}'_{D-p-2}(r)\, \widetilde{\epsilon}_{(D-p-2)}\, dr \ , \label{clo}
\eeq
where $b$ and $\tilde{b}$ are functions of $r$ and also the closed $r$-independent forms 
\beq 
h_{p+1}\ {\epsilon_{(p+1)}}  \ , \qquad \tilde{h}_{D-p-2}\ {\widetilde{\epsilon}_{(D-p-2)}} \ , \label{clo2}
\eeq 
where $h_{p+1}$ and $\tilde{h}_{D-p-2}$ are two constants. In components
\bea
{\cal H}_{p+2,\ \mu_1 \ldots \mu_{p+1} r} &=& \sqrt{-\gamma(x)}\ \epsilon_{\mu_1 \ldots \mu_{p+1}} \, b'{}_{p+1}(r) \ , \nonumber \\ {\cal H}_{D-p-1,\ i_1 \ldots i_{D-p-2}\,r} &=& \sqrt{\gamma(\xi)}\ \epsilon_{i_1 \ldots i_{D-p-2}} \, \tilde{b}'{}_{D-p-2}(r) \ ,  \label{profile_H}
\eea
and
\bea
{\cal H}_{p+1,\ \mu_1 \ldots \mu_{p+1}} &=& \sqrt{-\gamma(x)}\ \epsilon_{\mu_1 \ldots \mu_{p+1}} \, h_{p+1} \ , \nonumber \\
{\cal H}_{D-p-2,\ i_1 \ldots i_{D-p-2}} &=& \sqrt{\gamma(\xi)}\ \epsilon_{i_1 \ldots i_{D-p-2}} \, \tilde{h}_{D-p-2} \ . \label{profile_h}
\eea
We can now concentrate on the first member of each of the preceding couples, since the others are related to them by dualities.

The forms in eqs.~\eqref{clo} and \eqref{clo2} have degrees $p+2$, $D-p-1$, $p+1$ and $D-p-2$, and correspond to field strengths of forms of degrees $p+1$, $D-p-2$, $p$ and $D-p-3$. The two cases in eqs.~\eqref{clo2} are special, in that they are $r$--independent, and moreover the corresponding gauge fields are invariant under the isometry groups only up to gauge transformations. Therefore, they have to be treated with care, and we shall return to this issue in the following section, after eq.~\eqref{delta_h1}. Since all these forms are closed, the Bianchi identities are satisfied by the tensor profiles in eqs.~\eqref{clo} and \eqref{clo2}. Moreover, taking into account that
\bea
 {}^{*} \epsilon_{(p+1)}&=& e^{B-(p+1)A+(D-p-2)C} dr \wedge \widetilde{\epsilon}_{(D-p-2)} \ , \nonumber \\
{}^{*} \widetilde{\epsilon}_{(D-p-2)} &=& (-1)^{(p+2)(D-p-2)} \ e^{B+(p+1)A-(D-p-2)C} \epsilon_{(p+1)}\wedge dr \ .
\label{star_clo2}
\eea
one can see that the dynamical equations~\eqref{eqsbeta} are identically satisfied by the $r$--independent forms in eq.~\eqref{clo2}, for arbitrary $r$--dependent scalar profiles $\phi(r)$ and metric profiles in eq.~\eqref{metric_sym}. On the other hand, the two profiles in eq.~\eqref{clo} satisfy eqs.~\eqref{eqsbeta} if
\bea
b'_{p+1}(r) &=&   H_{p+2}\ e^{\,2\,\beta_p\phi + B +(p+1)A-(D-p-2)C} \, , \nonumber \\  b'_{D-p-2}(r) &=&  \widetilde{H}_{D-p-1}\, e^{\,2\,\beta_{D-p-3}\phi + B -(p+1)A+(D-p-2)C} \, , \label{h12}
\eea
where the factors are a pair of constants. 

Summarizing, in form language the four symmetric tensor profiles are described by
\bea
{\cal H}_{p+2} &=& H_{p+2}\ e^{\,2\,\beta_p\phi + B +(p+1)A-(D-p-2)C}\ \sqrt{-\gamma(x)} \ dx^0 \wedge \ldots \wedge dx^p \wedge dr \ , \nonumber \\
{\cal H}_{p+1} &=& h_{p+1}\, \sqrt{-\gamma(x)} \ dx^0 \wedge \ldots \wedge dx^p \ . \label{Hhforms1}
\eea
and by
\bea
{\cal H}_{D-p-1} &=& \widetilde{H}_{D-p-1}\ e^{\,2\,\beta_{D-p-3}\phi + B -(p+1)A+(D-p-2)C} \ dy^1 \wedge \ldots \wedge dy^{D-p-2} \wedge dr \ , \nonumber \\
{\cal H}_{D-p-2} &=& \tilde{h}_{D-p-2}\, \sqrt{\gamma(\xi)} \ dy^1 \wedge \ldots \wedge dy^{D-p-2} \ . \label{Hhforms2}
\eea

For brane profiles $k'$ would be equal to one, $r$ would be a radial coordinate and the internal space would be a sphere. The total ``electric'' charge of the first profile in eq.~\eqref{Hhforms1} would be finite and given by
\beq
{q}_e \ = \ H_{p+2}\, \Omega_{D-p-2} \ ,
\eeq
with $\Omega_{D-p-2}$ the volume of a unit internal sphere. The first profile would then be the configuration sourced by an electric $p$-brane, whose counterpart in Maxwell's theory is the Coulomb field of a point charge. On the other hand, with a sphere as internal space, the second profile in eq.~\eqref{Hhforms1}, which also respects the symmetry of the background, would be a uniform field in spacetime, which would result from uncharged open $p$-branes carrying, on their boundaries, opposite charges of one lower dimension, associated to $(p-1)$-(anti)branes. Its counterpart in the standard Maxwell's theory would be a uniform ``electric'' field resulting from a pair of opposite charges $q$ and $-q$ moved, in opposite directions, to a very large mutual distance $r_0$, in such a way the ratio $\frac{q}{r_0^2}$ remains finite. The second profile in eq.~\eqref{Hhforms2} would be the configuration sourced by a magnetic $p$-brane, with magnetic charge
\beq
{q}_m \ = \ h_{D-p-2}\, \Omega_{D-p-2} \ ,
\eeq
whose counterpart in the standard Maxwell theory would be the field of a magnetic monopole, while the first profile in eq.~\eqref{Hhforms2} is the dual of the second profile in eq.~\eqref{Hhforms1}. 

There are also some special tensor profiles that are relevant for type--IIB supergravity in ten dimensions. They lead to an interesting class of vacua, which will be dealt with at length in~\cite{int4d_vacuum}. A proper account of the contribution of these profiles requires a few additional comments, since the corresponding field strength is self--dual. To begin with, one can start from the solution of the self--duality condition, which reads
\beq
{\cal H}_{5} \ = \ \frac{H_{5}}{2 \sqrt{2}}\,\left(e^{B +4\,A\,-\,5\,C}\, {\epsilon}_{(4)}  \wedge dr  \,+\, \widetilde{\epsilon}_{(5)} \right) \ , \label{H5H}
\eeq
since $\beta=0$ in this case. For $k'=1$ this type of profile would be associated to a dyon. In a similar fashion, a second type of profile,
\beq
{\cal H}_{5} \ = \ \frac{h_{5}}{2 \sqrt{2}}\left(\epsilon_{(5)} \ + \ e^{ - 5 A + B + 4 C} \ dr \wedge \widetilde{\epsilon}_{(4)} \right) \ , \label{H5h}
\eeq
is the counterpart of eq.~\eqref{H5H} for the $h$ field strengths discussed above. 

\section{\sc Dynamical Action Principle and Equations of Motion} \label{sec:reduced_action}

In this section we derive a dynamical action principle inserting in eq.~\eqref{eqs4} the symmetric profiles that we have described.
To begin with, up to an overall factor that we shall leave out consistently, the metric of eq.~\eqref{metric_sym} and a symmetric scalar profile lead to the reduced action principle
\begin{align}
{\cal S} &= \frac{1}{2} \ \int dr \, \Big\{ \, e^{(p+1)A-B+(D-p-2)C}\Big[  p(p+1)\left(A'\right)^2 \ +\ (D-p-2)(D-p-3) \left( C'\right)^2 \nonumber \\ & - \ \frac{4\,(\phi')^2}{D-2} \ + \ 2(p+1)(D-p-2)A'\,C' \Big] \ + \, \frac{k}{\alpha'} \, p\,(p+1)\,e^{(p-1)A+B+(D-p-2)C} \nonumber \\ & - \, {T}_{D-1}\, e^{(p+1)A+B+(D-p-2)C+\gamma \phi}\ + \  \frac{k'}{\alpha'} \, (D-p-2)(D-p-3)\, e^{(p+1)A+B+(D-p-4)C} \Big\} \ ,
\end{align}
where 
\beq
T_{D-1} \ = \ \frac{\tau_{D-1}}{\alpha'} \ ,
\eeq
and $k$ and $k'$, as we have already stated, are the curvatures of the $(p+1)$--dimensional metric $\gamma_{\mu\nu}(x)$ and of the $(D-p-2)$--dimensional metric $\gamma_{mn}(\xi)$. From now on, for brevity, $T_{D-1}$ will be concisely denoted by $T$.

As we have seen, there are two independent options for the inclusion of symmetric tensor fluxes in the class of metric of eq.~\eqref{metric_sym}. The first one corresponds to the first profile in eq.~\eqref{clo} for a $(p+1)$--form gauge field, with $b'{}_{p+1}(r)$ given in eq.~\eqref{h12}, and contributes to the dynamical action principle the term
\beq
\Delta\,{\cal S}_{\cal H}^{(1)} \ = \ \frac{1}{4} \ \int dr\ e^{\,-\,2\,\beta_p\,\phi\,-\,(p+1)A \,-\, B \,+\, (D-p-2)C} \left(b'_{p+1}\right)^2 \ . \label{delta_h1}
\eeq
The second independent option corresponds to the first profile in eq.~\eqref{clo2} for a $p$--form gauge field, but is not described in these simple symmetrical terms, as we have stressed. Therefore we shall only include its contribution to the equations of motions, deducing it from eqs.~\eqref{eqsbeta} and \eqref{eqsnotlagbeta}, while also making use of eq.~\eqref{h2mn}.
With this proviso, and with different choices of $p$ and $\beta_p$, one can describe in this fashion, as we have anticipated, all symmetric ``electric'' and ``magnetic''  fluxes of interest.

One can now combine the different contributions described in Section~\ref{sec:symmetries}, aside from the one related to $h_{p+1}$, and the end result reads
\begin{align}
{\cal S} &= \frac{1}{2} \ \int dr \, \Big\{ \, e^{(p+1)A-B+(D-p-2)C}\Big[  p(p+1)\left(A'\right)^2 \ +\ (D-p-2)(D-p-3) \left( C'\right)^2 \nonumber \\ & - \ \frac{4\,(\phi')^2}{D-2} \ + \ 2(p+1)(D-p-2)A'\,C' \Big] \ + \, \frac{k}{\alpha'} \, p\,(p+1)\,e^{(p-1)A+B+(D-p-2)C} \nonumber \\ & - \, {T} \, e^{(p+1)A+B+(D-p-2)C+\gamma \phi}\ + \  \frac{k'}{\alpha'} \, (D-p-2)(D-p-3)\, e^{(p+1)A+B+(D-p-4)C}\nonumber \\
&+ \, \frac{1}{2}\ e^{\,-\,2\,\beta_p\,\phi\,-\,(p+1)A \,-\, B \, +\, (D-p-2)C} \ \left(b'_{p+1}\right)^2 
 \Big\} \label{reduced_action_part_int} \ .
\end{align}
In the resulting equations of motion, we shall also include shortly the contribution related to $h_{p+1}$, starting from eqs.~\eqref{eqsbeta} and~\eqref{eqsnotlagbeta}.  
To begin with, the first tensor profile of Section~\ref{sec:tensor_profiles} satisfies the simple equation
  \beq
 \left(e^{\,-\,2\,\beta_p\,\phi\,-\,(p+1)A \,-\, B \, +\, (D-p-2)C} \ b'_{p+1}\right)^\prime \ = \ 0 \ .
 \label{Eqb_red1}
\eeq
which is solved by
\beq
b'_{p+1}(r) \ = \   H_{p+2}\ e^{\,2\,\beta_p\phi + B +(p+1)A-(D-p-2)C}\ ,
\eeq
as we had seen more generally in eq.~\eqref{Hhforms1}. It is now convenient to work in the ``harmonic'' gauge 
\beq
 (p+1)A \ - \ B \ + (D-p-2) C \ = \ 0 \ , \label{F_gauge}
\eeq
which reduces the equations of motion for $A$, $C$ and $\phi$ to
 \bea
 A'' \!\!\!&=& - \ \frac{T}{(D-2)} \ e^{2\,B\,+\,\gamma\,\phi}\ +\ \frac{k\,p}{\alpha'}\ e^{2(B-A)} \label{EqA_red}\\
  &+& \!\!\frac{(D-p-3)}{2\,(D-2)} \ e^{2\,B\,+\,2\,\beta_p\,\phi\,-\,2(D-p-2)C } H_{p+2}^2 \,+\, \frac{(D-p-2)}{2\,(D-2)} \ e^{2\,B\,-\,2\,\beta_{p-1}\,\phi\,-\,2 (p+1) A} h_{p+1}^2  \nonumber \  ,  \\
C''\!\!\!&=& - \ \frac{T}{(D-2)} \ e^{2\,B\,+\,\gamma\,\phi}\ +\ \frac{k'(D-p-3)}{\alpha'}\ e^{2(B-C)} \label{EqC_red} \\
   &-& \frac{(p+1)}{2\,(D-2)}\ e^{2\,B\,+\,2\,\beta_p\,\phi\,-\,2(D-p-2)C }  H_{p+2}^2\,-\, \frac{p}{2\,(D-2)}\ e^{2\,B\,-\,2\,\beta_{p-1}\,\phi\,-\,2 (p+1) A} h_{p+1}^2 \ ,  \nonumber  \\
   \phi'' \!\!\!&=& \frac{T\,\gamma\,(D-2)}{8}\ e^{2\,B\,+\,\gamma\,\phi} \label{Eqphi_red} \\ &+& \frac{\beta_p\,(D-2)}{8}\ e^{2\,B\,+\,2\,\beta_p\,\phi\,-\,2(D-p-2)C } H_{p+2}^2 \,+\, \frac{\beta_{p-1}\,(D-2)}{8}\ e^{2\,B\,-\,2\,\beta_{p-1}\,\phi\,-\,2 (p+1) A} h_{p+1}^2\ . \nonumber
  \eea
  Here we have included the contributions related to $h_{p+1}$, and moreover the equation for $B$, which is usually called ``Hamiltonian constraint'', reads
 \begin{align}
&(p+1)A'[p\,A' \,+\, (D-p-2)C']\,+\, (D-p-2)C'[(D-p-3)C'+(p+1)A'] \nonumber \\
 &- \, \frac{4\,(\phi')^2}{D-2} \, + \, {T} \, e^{\, 2\,B\,+\,\gamma\,\phi}\, - \, \frac{k\,p(p+1)}{\alpha'}\ e^{2(B-A)}\, - \, \frac{k'(D-p-3)(D-p-2)}{\alpha'}\ e^{2(B-C)} \nonumber \\
 & + \, \frac{1}{2}\, e^{\,2\,\beta_p\,\phi\,+\,2\,B\,-\,2\,(D-p-2)\,C} \ H_{p+2}^2 \,-\, \frac{1}{2}\, e^{\,-\,2\,\beta_{p-1}\,\phi\,-\,2(p+1)A\,+\,2\,B} \ h_{p+1}^2  \,= \, 0 \ . \label{EqB_red}
 \end{align}
  
Notice that this system has an interesting discrete symmetry: its equations are left invariant by the redefinitions
\bea
&& \left[A,C,\,p,\,k,k'\right] \ \longleftrightarrow \ \left[C,A,\,D-p-3,\,k',k \right] \ , \nonumber \\
&& \left[H_{p+2}^2,\beta_p;h_{p+1}^2,\beta_{p-1} \right] \ \longleftrightarrow \ \left[-h_{p+1}^2,- \beta_{p-1};- H_{p+2}^2,-\beta_p  \right] \label{sym_AC}
\ , \eea
which can be regarded as implementing an ``electric-magnetic'' duality.
The simultaneous presence of $H_{p+2}$ and $h_{p'+1}$ profiles is only relevant in one special case, for type IIA, with one of them of NS-NS type and the other of RR type.  
In the following, we shall focus on solutions with $k=k'=0$, leaving the study of the other types of configurations to~\cite{deformed_branes}.

As we have anticipated, two special cases, related to the type--IIB string, must be treated separately, since they involve fluxes of the self--dual five--form field strength, for which we refer the reader to eqs.~\eqref{H5H} and \eqref{H5h}.
The complete equations of motion for the first case are
\beq
R_{MN}  \ = \  \frac{1}{24}\ \left({\cal H}_{5}^2\right)_{M N} \ + \ \frac{1}{2}\, \partial_M\,\phi \,\partial_N\,\phi \ ,
\eeq
and their reduced form for the class of metrics of interest in the ``harmonic'' gauge $B=4 A + 5 C$ and for the symmetric $H_5$ profile of eq.~\eqref{H5H} reads
\bea
A'' &=&  \frac{H_5^2}{8}\ e^{8 A} \ , \nonumber \\
C'' &=&  - \ \frac{H_5^2}{8}\ e^{8 A}\ , \nonumber \\
\phi'' &=& 0 \ . \label{eqABC_sdual}
\eea
The corresponding Hamiltonian constraint is
\beq
3\left(A'\right)^2 \ + \ 10\, A'\, C' \ + \ 5 \left(C'\right)^2 \ = \ \frac{1}{8}\, \left(\phi'\right)^2 \ - \ \frac{H_5^2}{16} \ e^{8 A} \ . \label{ham_sdual}
\eeq
The counterpart of these results for the $h_{p+1}$--fluxes corresponds to $p=4$, and in this case
\bea
A'' &=& \frac{h_5^2}{8}\ e^{8 C} \ , \nonumber \\
C'' &=&  - \ \frac{h_5^2}{8}\ e^{8 C}\ , \nonumber \\
\phi'' &=& 0 \ .  \label{eqABC_sdual_h}
\eea
while the Hamiltonian constraint becomes
\beq
5\left(A'\right)^2 \ + \ 10\, A'\, C' \ + \ 3 \left(C'\right)^2 \ = \ \frac{1}{8}\, \left(\phi'\right)^2 \ + \ \frac{h_5^2}{16} \ e^{8 C} \ . \label{ham_sdual_h}
\eeq

\section{\sc Supersymmetric Vacua with Fluxes} \label{sec:killing_spinors}

In this section we determine which backgrounds of the general form of eq.~\eqref{metric_sym} with $k=k'=0$, with corresponding symmetric profiles as in eq.~\eqref{Hhforms1}, are supersymmetric. These are special solutions of eqs.~\eqref{EqA_red}--\eqref{EqB_red}, and in a particular case of eqs.~\eqref{eqABC_sdual} and \eqref{ham_sdual}, which maintain part of the original ten--dimensional supersymmetry, and we shall obtain them solving the first--order Killing equations. The corresponding solutions for $k'=1$ are the well-known BPS brane profiles (for a review, see~\cite{duff} or Polchinski's book in~\cite{strings}). We begin from the case of eleven--dimensional supergravity~\cite{cjs}, which is simpler since it does not contain a dilaton field.

\subsection{\sc Supersymmetric Vacua of Eleven--Dimensional Supergravity}
In this case the starting point is provided by the supersymmetry transformations of eleven--dimensional gravitino~\cite{cjs},
\beq{}{}{}
\delta\,\psi_M \,=\, D_M\,\epsilon \ + \ \frac{1}{288} \left( \Gamma_{M}{}^{N_1 N_2 N_3 N_4} \,-\,8\, \delta_M{}^{N_1}\,\Gamma^{N_2 N_3 N_4}\right) {\cal H}_{N_1 N_2 N_3 N_4}\,\epsilon \ ,
\eeq
where
\bea{}{}{}
D_\mu\,\epsilon &=& \partial_\mu\,\epsilon \ + \ \frac{1}{2}\, A' \,e^{A-B} \gamma_\mu\,\gamma_r \,\epsilon \ , \nonumber \\
D_r\,\epsilon &=& \partial_r\,\epsilon \ , \nonumber \\
D_m\,\epsilon &=& \partial_m\,\epsilon \ + \ \frac{1}{2}\,C' \,e^{C-B} \gamma_m\,\gamma_r \,\epsilon \ .
\eea
Following the discussion in Section~\ref{sec:tensor_profiles}, one can see that there are two $H$-type field strengths of interest, with $p=2,5$, and correspondingly two possible cases.
\begin{itemize}
\item{For $p=2$ there is an internal $T^7$, and the $H$-form in the first of eqs.~\eqref{Hhforms1}, in the harmonic gauge and with no dilaton, reads }
\beq{}{}{}
{\cal H}_4 \ = \ H_4 \, e^{6A}\,\epsilon_3\,\wedge\,  dr \ .
\eeq
Moreover, in this symmetric profile the three groups of supersymmetry transformations become
\bea{}{}{}
\delta\,\psi_\mu &=& \partial_\mu\,\epsilon \ + \ \frac{1}{2}\,\gamma_\mu\,\gamma_r\left[A' \,e^{A-B} \,+\,\frac{H_4}{3}\,e^{4A-B} \,\gamma \right] \epsilon \ , \nonumber \\
\delta\,\psi_r &=& \partial_r\,\epsilon \,+\,\frac{H_4}{6}\,e^{3A}\,\gamma\,\epsilon \ , \label{delta11_2} \\
\delta\,\psi_m &=& \partial_m\,\epsilon \,+\, \frac{1}{2}\,\gamma_m\,\gamma_r\left[C' \,e^{C-B} \ - \ \frac{H_4}{6}\, e^{3A+C-B}\,\gamma\right]\epsilon\ , \nonumber
\eea
where the matrix $\gamma$, defined by
\beq{}{}{}
\gamma \ = \ \frac{1}{6}\, \epsilon_{\mu\nu\rho} \,\gamma^{\mu\nu\rho} \ ,
\eeq
satisfies
\beq{}{}{}
\gamma^2 \ = \ 1\ , \qquad \gamma_\mu\,\gamma \ = \ - \ \gamma\,\gamma_\mu \ = \ \frac{1}{2}\,\epsilon_{\mu\nu\rho}\,\gamma^{\nu\rho} \ .
\eeq
One can now solve the conditions
\beq{}{}{}{}
\delta\,\psi_M \ = \ 0  \ ,
\eeq
making use of their explicit form in eqs.~\eqref{delta11_2}, which leads to
\bea{}{}{}
&&\gamma\,\epsilon \ = \ - \ \epsilon \ , \qquad \partial_\mu\,\epsilon \ = \ 0 \ , \qquad \partial_m\,\epsilon \ = \ 0 \ , \nonumber \\
&& \left(e^{\,-\,3\,A}\right)' \ = \ H_4\,r \ , \qquad C'\ = \ - \ \frac{1}{2}\, A'\ .
\eea
Notice that, if $H_4=0$, flat space is the only solution. Similar considerations apply to all the following cases.

Up to a translation of $r$ and a rescaling of the $y$ coordinates, the metric reads
\bea
ds^2 &=& \left(\left|H_4\right|r\right)^{\,-\, \frac{2}{3}}\, dx^2 \,+\,  \left(\left|H_4\right|r\right)^{\,\frac{1}{3}}\left(dr^2 \,+\, dy^2\right) \ . \label{solp2}
\eea
Finally, the condition $\delta\,\psi_r=0$ gives the differential equation
\beq{}{}{}{}
\partial_r\,\epsilon \ + \ \frac{1}{6\,r}\,\epsilon \ = \ 0 \ , 
\eeq
which determines the $r$--dependent spinor profile and is solved by
\beq
\epsilon(r) \ = \ \frac{\epsilon_0}{\left(\left|H_4\,r\right|\right)^\frac{1}{6}} \ ,
\eeq
with $\epsilon_0$ a constant spinor subject to the condition
\beq{}{}{}{}
\gamma\,\epsilon_0 = - \ \epsilon_0 \ .
\eeq
The preceding results also determine the tensor profile
\beq{}{}{}
{\cal H}_4 \ = \ H_4 \ \frac{\epsilon_3\,\wedge\,  dr}{\left(\left|H_4\right| r\right)^2} \ ,
\eeq
whose dual is
\beq{}{}{}
\star {\cal H}_4 \ = \ H_4 \ {\widetilde{\epsilon}_7} \ .
\eeq
\item{In the $p=5$ case there is an internal $T^4$, and the dual of the seven--form $H$-field strength is} 
\beq{}{}{}
{\cal H}_4 \ = \ H_7 \,\widetilde{\epsilon}_4 \ .
\eeq
The three groups of supersymmetry transformations read
\bea{}{}{}
\delta\,\psi_\mu &=& \partial_\mu\,\epsilon \ + \ \frac{1}{2}\,\gamma_\mu\,\gamma_r\left[A' \,e^{A-B} \,+\,\frac{H_7}{6}\,e^{A-4C} \,\gamma_r\,\gamma \right] \epsilon \ , \nonumber \\
\delta\,\psi_r &=& \partial_r\,\epsilon \,+\,\frac{H_7}{12}\,e^{B-4C}\,\gamma_r\,\gamma\,\epsilon \ , \label{delta11_5} \\
\delta\,\psi_m &=& \partial_m\,\epsilon \,+\, \frac{1}{2}\,\gamma_m\,\gamma_r\left[C' \,e^{C-B} \ - \ \frac{H_7}{3}\, e^{-3C}\,\gamma_r\,\gamma\right]\epsilon\ . \nonumber
\eea
Now, defining $\gamma$ as
\beq{}{}{}{}
\gamma \ = \ \gamma^7\,\gamma^8\,\gamma^9\,\gamma^{10} \ ,
\eeq
the Killing spinor equations require that $\epsilon$ be an eigenvector of $\gamma\,\gamma_r$, and one can make the choice
\beq{}{}{}{}
\gamma_r\,\gamma\,\epsilon \ = \ \epsilon \ . \label{gammargamma}
\eeq
This is possible since
\beq{}{}{}{}
\left(\gamma_r \, \gamma\right)^2 \ = \ 1 \ ,
\eeq
and then
\bea
ds^2 &=& \left(\left|H_7\right|r\right)^{\,-\, \frac{1}{3}}\, dx^2 \,+\,  \left(\left|H_7\right|r\right)^{\,\frac{2}{3}}\left(dr^2 \,+\, dy^2\right) \ , \nonumber \\
\epsilon(r) &=& \frac{1}{\left( \left|H_7\right|r\right)^\frac{1}{12}} \, \epsilon_0 \ , \label{solp5}
\eea
with $\epsilon_0$ a constant spinor subject to the projection~\eqref{gammargamma}.
\end{itemize}

Summarizing, we have seen that eleven--dimensional supergravity admits, in addition to flat space, which preserves 32 supersymmetries, other backgrounds with $H$-fluxes that preserve 16 supersymmetries. They are in one-to-one correspondence with BPS brane solutions, and the difference is that spheres in the transverse space are replaced by flat space or tori. In the latter case they describe compactifications to three or six dimensions. Following similar steps, one can see that there are no corresponding solutions with $h$-fluxes.

We can now turn to discuss ten--dimensional supersymmetric strings, referring explicitly to type IIA and IIB models, although part of the latter options clearly apply to type I. One must distinguish further two cases, according to whether ${\cal H}_{p+2}$ is an RR or NS-NS form, and the self--dual five-form field strength of type IIB will be treated separately.

\subsection{\sc RR $H$--Forms in Ten Dimensions}
Let us begin our discussion from the case of RR forms. We can now let $D=10$, while also recalling that $p$ is even for type IIA and odd for type IIB. We work in the string frame, in order to rely directly on the setup of~\cite{bkrvp}, while also leaving aside momentarily the $p=3$ case, so that the metric has the form of eq.~\eqref{metric_sym_intro} with $A$, $B$ and $C$ replaced by
\beq{}{}
A_s\,=\,A\,+\,\frac{\phi}{4} \ , \quad B_s\,=\,B\,+\,\frac{\phi}{4} \ , \quad C_s\,=\,C\,+\,\frac{\phi}{4} \ , \ \label{Einststring}
\eeq{}
and consequently the harmonic gauge condition~\eqref{F_gauge} becomes
\beq{}{}
B_s \ = \ (p+1)\,A_s \ + \ (8-p)\,C_s \ - \ 2\,\phi \ . \label{harm_S}
\eeq
In the string frame, as we stressed in Section~\ref{sec:tensor_profiles}, $\beta_S=0$, and taking into account the gauge condition, one can translate the first of eqs.~\eqref{Hhforms1}, into
\beq
{\cal H}_{(p+2)} \,=\, H_{p+2}\,e^{B_s+(p+1)A_s-(8-p)C_s}\,dx^0 \wedge \ldots \wedge dx^p \wedge dr \ . \label{form_p}
\eeq
Here all $\gamma$ matrices are flat,
so that taking into account the pairs of ``dual'' contributions and using
\beq
\frac{\slashed{\cal H}_{(p+2)}}{(p+2)!} \,=\, H_{p+2}\ e^{\,-\,(8-p)C_s} \ \gamma^{0\ldots p}\,\gamma^r \ ,
\eeq
the supersymmetry transformations of the Fermi fields taken from~\cite{bkrvp} finally read
\bea
\delta\,\lambda &=& e^{-B_s}\left(\gamma_r \,\phi' \,\epsilon \ + \ H_{p+2}\ e^{\,-\,\phi\, +\, (p+1)A_s}\ (-1)^p \ \frac{ (3-p)}{4} \ \gamma^{0\ldots p} \,\gamma_r \,{\cal P}_{\frac{p}{2}+1} \epsilon\right) \ , \nonumber \\
\delta\,\psi_r &=& \partial_r \, \epsilon \, + \, \frac{H_{p+2}}{8} \ e^{\,-\,\phi\, +\, (p+1)A_s}\ \gamma^{0\ldots p} \,{\cal P}_{\frac{p}{2}+1} \epsilon \ ,
\label{killing_spinors2} \\
\delta\,\psi_\mu &=& \partial_\mu \, \epsilon \,+\, \frac{1}{2}\ \gamma_\mu \gamma_r\ e^{-pA_s-(8-p)C_s+2\phi}\, A_s'\,\epsilon \, + \, \frac{H_{p+2}}{8} \ e^{\,\phi\, +\, A_s\,-\,(8-p)C_s}\  \gamma^{0\ldots p} \gamma_r\,\gamma_\mu\,{\cal P}_{\frac{p}{2}+1} \epsilon \ ,
\nonumber \\
\delta\,\psi_m &=& \partial_m \, \epsilon \,+\, \frac{1}{2}\ \gamma_m \gamma_r\ e^{2\phi-(p+1)A_s-(7-p)C_s}\, C_s'\,\epsilon \, + \, \frac{H_{p+2}}{8} \ e^{\,\phi\,-\,(7-p)C_s}\  \gamma^{0\ldots p} \gamma_r\,\gamma_m\,{\cal P}_{\frac{p}{2}+1} \epsilon \ , \nonumber
\eea
after making use of the gauge condition~\eqref{harm_S}.
Here ${\cal P}_{\frac{p}{2}+1}$~\cite{bkrvp} has the following form:
\begin{enumerate}
    \item for type IIA \  ${\cal P}_{\frac{p}{2}+1} = \left(\gamma_{11}\right)^{\frac{p}{2}+1}$, with $p$ even;
    \item for type IIB \  ${\cal P}_{\frac{p}{2}+1} = \sigma_1$ if  $p=1,5$, and ${\cal P}_{\frac{p}{2}+1} = i \sigma_2$ in the remaining cases.
\end{enumerate}

For $p \neq 3$, one can conveniently start from the first of eqs.~\eqref{killing_spinors2}, demanding that
\bea
\delta\,\lambda &=& e^{-B}\gamma_r \,\phi' \left(\epsilon \ - \ \frac{H_{p+2}}{\phi'}\ e^{\,-\,\phi\, +\, (p+1)A_s}\ \frac{ (3-p)}{4} \ \gamma^{0\ldots p} \,{\cal P}_{\frac{p}{2}+1} \epsilon\right) \ = \ 0 \ . \label{delta_lambda}
\eea
This equation involves a projection on $\epsilon$ provided
\beq
\phi'\ = \ \pm \ \frac{(3-p)}{4} \ {H_{p+2}}\ e^{\,(p+1)A_s-\phi}\ , \label{cons_phi}
\eeq
since then it reduces to
\beq
\left(1 \ \mp \ \gamma^{0\ldots p} \,{\cal P}_{\frac{p}{2}+1} \right) \epsilon \ = \ 0 \ , \label{susy_proj}
\eeq
which halves the number of supersymmetries with respect to those present in the ten--dimensional Minkowski vacuum.
One can see that in all cases, for both IIA and IIB, 
\beq
\left(\gamma^{0\ldots p} \,{\cal P}_{\frac{p}{2}+1}\right)^2 \ = \ 1 \ ,
\eeq
and therefore the preceding condition is indeed a consistent projection. For $p \neq 3$ the condition $\delta\,\psi_r=0$ can be turned into
\beq
\left(\partial_r \ + \ \frac{1}{2(3-p)}\ \phi' \right) \epsilon \ = \ 0 \ ,
\eeq
whose solution is 
\beq
\epsilon \ = \ e^{\,-\,\frac{\phi}{2(3-p)}} \ \epsilon_0(x,y) \ , \label{epsilon_epsilon0}
\eeq
while the condition $\delta\,\psi_\mu=0$ reads
\beq
\partial_\mu \epsilon \ + \ \frac{1}{2}\ \gamma_\mu\,\gamma_r\, e^{-p A_s-(8-p)C_s+2\phi} \left( A_s' \ + \ \frac{\phi'}{(3-p)} \right) \epsilon \ = \ 0 \ .
\eeq
Since the projection anticommutes with $\gamma_\nu\,\gamma_r$, one thus gets the two conditions
\bea
A_s' \ + \ \frac{\phi'}{(3-p)} &=& 0 \ , \nonumber \\
\partial_\mu\, \epsilon &=& 0 \ . \label{eqsAphi}
\eea
In a similar fashion, from $\delta\,\psi_m=0$ one can deduce that
\beq
\partial_m \epsilon \ + \ \frac{1}{2}\ \gamma_m\,\gamma_r\, e^{2\phi-(p+1)A_s-(7-p)C_s}\left( C_s' \ - \ \frac{\phi'}{(3-p)} \right) \epsilon \ = \ 0 \ . \label{deltapsii}
\eeq
Making use of eq.~\eqref{epsilon_epsilon0}, one can now replace $\epsilon$ with $\epsilon_0$, and then all residual $r$--dependence must disappear, so that
\beq
e^{2\phi-(p+1)A_s-(7-p)C_s} \left( C_s' \ - \ \frac{\phi'}{(3-p)} \right)  \ = \ -\ 2\,\sigma \ ,
\eeq
with $\sigma$ a constant. Eq.~\eqref{deltapsii} thus reduces to
\beq
\partial_m \epsilon \ = \ \sigma\ \gamma_m\,\gamma_r\,  \epsilon \ ,
\eeq
which implies
\beq
\partial_m\,\partial_n\,\epsilon \ = \ - \ \sigma^2 \, \gamma_m\,\gamma_n\, \epsilon \ .
\eeq
Since the partial derivatives commute, the only possible choice is $\sigma=0$, so that finally
\beq
C_s' \ - \ \frac{\phi'}{(3-p)} \ = \ 0  \ . \label{eqsCphi}
\eeq
Consequently, $\epsilon_0$ is also independent of the internal coordinates.

Summarizing, we have found that
\beq
\epsilon \ = \  e^{\,-\,\frac{\phi}{2(3-p)}} \ \epsilon_0 \ , 
\eeq
where $\epsilon_0$ is a constant spinor subject to the projection~\eqref{susy_proj}, and integrating eqs.~\eqref{eqsAphi} and \eqref{eqsCphi}
\beq
A_s  \ = \ - \ \frac{\phi}{(3-p)} \ + \ a_s \ , \qquad C_s  \ = \ \frac{\phi}{(3-p)} \ + \ c_s \ ,
\eeq
with $a_s$ and $c_s$ two additive constants, so that
\beq{}{}
e^{2A_s} \,=\, e^{\,-\,\frac{2\phi}{3-p} \,+\, 2\,a_s} \ , \qquad e^{2C_s} \,=\, e^{\,\frac{2\phi}{3-p} \,+\, 2\,c_s} \ , \qquad e^{2B_s} \,=\, e^{\,\frac{2\,\phi}{3-p} \,+\, 2(p+1)a_s+2(8-p)c_s}\ .
\eeq{}
Now eq.~\eqref{cons_phi} becomes
\beq
\phi' \ = \ \pm\ \frac{3-p}{4}\ H_{p+2}\ e^{\,-\,\frac{4\phi}{(3-p)}+(p+1)a_s} \ ,
\eeq
so that for $p \neq 3$ the string--frame solution reads
\beq
e^\phi \,=\, e^{\phi_s}\ \left[|H_{p+2}|\,r\right]^{\,\frac{3-p}{4}} \ ,
\eeq{}
in the region $r>0$, where
\beq{}{}
e^{\phi_s} \ = \ e^{\frac{(p+1)(3-p)}{4}\,a_s} \ .
\eeq{}
Consequently, after rescaling the $y$ coordinates, for $p \neq 3$ the string--frame metric and the form profile read
\bea
ds^2 &=& e^{\,\frac{3-p}{2}\,a_s}\,\left[|H_{p+2}|\,r\right]^{\,-\,\frac{1}{2}} dx^2 \ + \ \left[|H_{p+2}|\,r\right]^{\,\frac{1}{2}} \left(e^{\,2\,b_s}\ dr^2 \ + \  e^{\,2\,d_s}\ dy^2\right) \ , \nonumber \\
{\cal H}_{(p+2)} &=& \frac{H_{p+2}}{\left|H_{p+2}\,r\right|^{2}}\,dx^0 \wedge \ldots \wedge dx^p \wedge dr \ , 
\label{solnot3}
\eea
taking also eq.~\eqref{form_p} into account, where
\bea{}{}
2\,b_s &=& \frac{5}{2}\,(p+1)\,a_s + 2(8-p)c_s \ , \nonumber \\
2\,d_s &=& \frac{p+1}{2}\,a_s + 2 c_s \ .
\eea

More conveniently, one can rescale the $x$, $y$ and $r$ coordinates, in such a way that the three vacuum profiles become
\bea
ds^2 &=& \left[|H_{p+2}|\,r\right]^{\,-\,\frac{1}{2}} dx^2 \ + \ \left[|H_{p+2}|\,r\right]^{\,\frac{1}{2}} \left(dr^2 \ + \  dy^2\right) \ , \nonumber \\
{\cal H}_{(p+2)} &=& e^{\,\frac{(3-p)(8-p)}{5}\,c_s}\,\frac{H_{p+2}}{\left|H_{p+2}\,r\right|^{2}}\,dx^0 \wedge \ldots \wedge dx^p \wedge dr \ , \nonumber \\
e^\phi &=&  e^{\,-\,\frac{(3-p)(8-p)}{5}\,c_s}\,\left[|H_{p+2}|\,r\right]^{\,\frac{3-p}{4}} \ ,
\label{solnot3r}
\eea
while the Killing spinor becomes
\beq
\epsilon \ = \ {e^{\,-\,\frac{\phi}{2(3-p)}}} \ \epsilon_0 \ , \qquad \left(1 \ \mp \ \gamma^{0\ldots p} \,{\cal P}_{\frac{p}{2}+1} \right) \epsilon_0 \ = \ 0 \ , \label{killingnot3}
\eeq
with $\epsilon_0$ a constant spinor. Notice also that
\beq{}{}
\star {\cal H}_{(p+2)} \,=\, e^{\,\frac{(3-p)(8-p)}{5}\,c_s}\,{H_{p+2}}\ dy^1 \wedge \ldots \wedge dy^{8-p} \ . 
\eeq
The corresponding form of the metric in the Einstein frame is determined by eqs.~\eqref{solnot3} or \eqref{solnot3r}, using eq.~\eqref{Einststring}, and reads
\beq{}{}
ds^2 \,=\, e^{\,\frac{(3-p)(8-p)}{10}\,c_s}\,\left[ \left[|H_{p+2}|\,r\right]^{\,\frac{p-7}{8}} dx^2 \ + \ \left[|H_{p+2}|\,r\right]^{\,\frac{p+1}{8}} \left(dr^2 \ + \  dy^2\right) \right] \ . \label{metricsusypE}
\eeq{}

The special case $p=3$ is discussed at length in~\cite{int4d_vacuum}, and here we can just summarize the results,
\bea
ds^2 &=&  \left[\frac{\sqrt{2}}{|H_5|\,r}\right]^\frac{1}{2}  \, dx^2 \, + \,  \left[\frac{|H_5|\,r}{\sqrt{2}}\right]^\frac{1}{2} \left(dr^2 \,+\,   dy^2\right) \ , \nonumber \\
{\cal H}_{5} &=& \frac{H_{5}}{2\sqrt{2}}\left[ \left[\frac{\sqrt{2}}{|H_5|\,r}\right]^{2} \, dx^0 \wedge \ldots \wedge dx^3 \wedge dr  \,+\, dy^1 \wedge \ldots \wedge dy^5 \right]\ , \nonumber \\
\phi &=& \mathrm{const} \ .
\label{killing_data}
\eea
This background is compatible with the existence of the Killing spinor
\beq
\epsilon \ = \  \left[\frac{\sqrt{2}}{|H_5|\,r}\right]^\frac{1}{8}\, \epsilon_{0} \ , \label{killingeps}
\eeq
where $\epsilon_0$ is a constant spinor subject to the condition
\beq
\gamma^{0\ldots 3}\,i\,\sigma_2\, \epsilon_0 \ = \ \sign\left(H_5\right) \epsilon_0 \ , \label{Lambdasusy}
\eeq
The reader should note the self--dual nature of the form and the absence of an overall constant.

\subsection{\sc NS--NS $H$--Forms in Ten Dimensions}

In this case the values of the constant $\beta$ of Section~\ref{sec:symmetries} are different, and lead to a different set of supersymmetry transformations. Here we shall elaborate separately on the two cases of interest, $p=1$ and $p=5$.

\subsubsection{\sc The $p=1$ case}

Let us begin by considering the NS-NS three--form, for which the supersymmetry transformations become
\bea
\delta\,\lambda &\equiv& e^{-B_s}\,\gamma_r \left( \phi'\, +\,\frac{1}{2}\, H_{3}\,e^{\,2\,\phi\,+\,B_s\,-\,7\,C_s}\,\gamma^{01}\,{\cal P}\right)\epsilon \ = \ 0 \ , \nonumber \\
\delta\,\psi_r &\equiv& \left(\partial_r  \,+\,  \frac{1}{4}\ H_{3}\, e^{\,2\,\phi\,+\,B_s\,-\,7\,C_s}\,\gamma^{01}\,{\cal P} \right) \epsilon  \ = \ 0 \ , \nonumber \\
\delta\,\psi_\mu &\equiv& \left(\partial_\mu  \,+\, \frac{1}{2}\ \gamma_\mu \gamma_r\ e^{A_s-B_s}\, A_s'\,+\,  \frac{1}{4}\ H_{3}\, \epsilon_{\mu\nu}\, e^{\,2\,\phi\,+\,A_s\,-\,7\,C_s}\,\gamma^{\nu}\,\gamma^r\,{\cal P}\right)\epsilon \ = \ 0 \ , \nonumber \\
\delta\,\psi_m &\equiv& \left(\partial_m \,+\, \frac{1}{2}\ \gamma_m \gamma_r\ e^{C_s-B_s}\, C_s'\right) \epsilon  \ = \ 0 \ ,
\label{killing_spinors_NS}
\eea
taking into account that for the NS two--form $\beta_p=1$ in the string frame, and making use of eq.~\eqref{Hhforms1}.
Here ${\cal P}$~\cite{bkrvp} has the following form:
\begin{enumerate}
    \item For type IIA: \  ${\cal P} = \gamma_{11}$;
    \item For type IIB: \  ${\cal P} = -\, \sigma_3$.
\end{enumerate}
To begin with, let us note that, in all cases,
\beq
\left(\gamma^{01}\,{\cal P}\right)^2 \ = \ 1 \ ,
\eeq
so that the variation of $\lambda$ tells us that
\beq
\gamma^{01}\,{\cal P}\, \epsilon \ = \ \sigma\, \epsilon \ , \label{proj_ns}
\eeq
with $\sigma = \pm 1$, and then
\beq
\phi'\  =\  - \ \frac{\sigma\,H_{3}}{2}\,e^{\,2\,\phi\,+\,B_s\,-\,7\,C_s} \ .
\eeq
The second equation is thus solved by
\beq
\epsilon \ =\ e^{\,\frac{\phi}{2}}\, \epsilon_0(x,y) \ ,
\eeq
and then, as in previous cases, the last equation implies that 
\beq
C_s'\ = \ 0 \ , \qquad \partial_i\, \epsilon_0 \ = \ 0 \ .
\eeq
Now the third equation reduces to
\beq
\left(\partial_\mu  \,+\, \frac{1}{2}\ \gamma_\mu \gamma_r\ e^{A_s-B_s}\left( A_s'\,-\, \phi'\right)  \right)\epsilon  \ = \ 0 \ ,
\eeq
using $\epsilon_{\mu\nu}\,\gamma^\nu \ = \ \gamma_\mu\,\gamma^{01}$, and the integrability of these conditions translates, as in previous cases, into
\beq
\partial_\mu\,\epsilon_0 \ = \ 0 \ , \qquad A_s'\ = \phi' \ ,
\eeq
so that $A_s = \phi+a_s$, where $a_s$ is a constant. The resulting equation for $\phi$ is
\beq
\phi'\ = \ - \ \frac{\sigma\,H_3}{2} \ e^{\,2\left(\phi\,+\,a_s\right)} \ ,
\eeq
which is solved by
\beq
e^{\,2\,\phi} \ = \ \frac{e^{\,-\,2\,a_s}}{\left(\sigma\,H_3\,r\right)} \ ,
\eeq
up to a translation of $r$. The gauge condition~\eqref{harm_S} determines now
\beq
B_s \ = \ 2\, a_s \ +\ 7\,C_s \ ,
\eeq
and one is thus finally led to the string--frame results
\bea
ds^2 &=& \frac{dx^2}{r} \ + \ dr^2 \ + \ dy^2 \ , \nonumber \\
e^\phi &=& \frac{e^{\,\phi_0}}{r^\frac{1}{2}} \ , \nonumber \\
\epsilon &=& \frac{1}{r^\frac{1}{4}}\, \epsilon_0 \ , \nonumber \\
{\cal H}_3 &=& \sigma\,\frac{dx^0 \wedge dx^1 \wedge dr}{r^2} \ .
\eea
Here $\sigma$ is a sign, and this simple result is reached after rescaling the $x$ and $y$ coordinates and $\epsilon_0$, which is a constant spinor subject to the projection~\eqref{proj_ns}, and after some redefinitions of the constants. In the Einstein frame, the metric becomes
\beq{}{}{}{}
ds^2 \ = \  \frac{dx^2}{r^\frac{3}{4}} \ + \ r^\frac{1}{4}\left(dr^2 \ + \ dy^2\right) \ .
\eeq
\subsubsection{\sc The $p=5$ case} 

Recasting now the seven--form field strength in terms of a dual three-form, as
\beq
{\widetilde{\cal H}}_3 \ = \ H_7 \ dy^1 \wedge dy^2 \wedge dy^3 \ ,
\eeq
the relevant portions of the supersymmetry transformations of eqs.~\eqref{killing_spinors2} become
\bea
\delta\,\lambda &\equiv& e^{-B_s}\,\gamma_r \left( \phi'\, +\,\frac{1}{2}\, H_{7}\,e^{\,B_s\,-\,3\,C_s}\,\gamma_r\,\gamma^{123}\,{\cal P}\right)\epsilon \ = \ 0 \ , \nonumber \\
\delta\,\psi_r &\equiv& \partial_r  \, \epsilon  \ = \ 0 \ , \nonumber \\
\delta\,\psi_\mu &\equiv& \left(\partial_\mu  \,+\, \frac{1}{2}\ \gamma_\mu \gamma_r\ e^{A_s-B_s}\, A_s'\right)\epsilon \ = \ 0 \ , \nonumber \\
\delta\,\psi_m &\equiv& \left(\partial_m \,+\, \frac{1}{2}\ \gamma_m \gamma_r\ e^{C_s-B_s}\, C_s' \,+\, \frac{1}{8}\, H_7\,\epsilon_{mnp}\,\gamma^{np}\,e^{\,-\,2\,C_s}\,{\cal P}\right) \epsilon  \ = \ 0 \ ,
\label{killing_spinors_NS_dual_p5}
\eea
and imply that now $\epsilon$ is a constant spinor, say $\epsilon_0$, with
\bea
&& \gamma_r\,\gamma^{123}\,{\cal P}\,\epsilon_0 \ = \ \sigma\,\epsilon_0 \ , \nonumber \\
&& A_s'\ = \ 0 \ , \nonumber \\
&& C_s' \ = \ \phi' \ , \nonumber \\
&& \phi'\ = \ - \ \frac{H_7\,\sigma}{2}\ e^{\,6 A_s\,-\,2\,\phi} \ ,
\eea
where $\sigma = \pm 1$. Consequently, working in the region $r>0$, one can choose
\beq
e^\phi \ = \ \left(\left|H_7\right|\,r \right)^\frac{1}{2} \ ,
\eeq
while the string--frame metric is
\beq
ds^2 \ =\ dx^2 \ + \ \left|H_7\right| \,r \left( dr^2 \ + \ dy^2 \right) \ ,
\eeq
up to rescalings of the $x$ and $y$ coordinates. With some redefinitions as above, these results can also be presented in the form
\bea
ds^2 &=& dx^2 \ + \ \left|r\right| \left( dr^2 \ + \ dy^2 \right) \ , \nonumber \\
e^\phi &=& \left|r\right|^\frac{1}{2}\ e^{\phi_0} \ , \nonumber \\
{\widetilde{\cal H}}_3 &=& dy^1 \wedge dy^2 \wedge dy^3 \ .
\eea
In the Einstein frame, the metric becomes
\beq{}{}{}{}
ds^2 \ = \  \frac{dx^2}{r^\frac{1}{4}} \ + \ r^\frac{3}{4}\left(dr^2 \ + \ dy^2\right) \ .
\eeq

\subsection{\sc $h$--Forms}

Let us begin from RR forms, where our starting point is
\beq
\frac{\slashed{\cal H}_{(p+1)}}{(p+1)!}  \ = \ h_{p+1}\ e^{\,-(p+1)A_s} \ \gamma^{0\ldots p} \ ,
\eeq
where the form profile was given in eq.~\eqref{Hhforms1},
and the supersymmetry transformations include
\beq
\delta\,\lambda \ = \  e^{-B_s}\gamma_r \left(\phi' \ + \ h_{p+1}\ e^{\,\phi\,+\,B_s\,-\,(p+1)A_s}\ \frac{ (4-p)}{4} \ \gamma^{0\ldots p} \,\gamma_r\,{\cal P}_{\frac{p+1}{2}} \right)\epsilon \ .
\label{killing_spinorsh}
\eeq
However, in contrast with the preceding cases, now
\beq
\left( \gamma^{0\ldots p} \,\gamma_r\,{\cal P}_{\frac{p+1}{2}} \right)^2 \ = \ - \ 1 \ ,
\eeq
and consequently the spinor would have to satisfy
\beq
\gamma^{0\ldots p} \,\gamma_r\,{\cal P}_{\frac{p+1}{2}} \ \epsilon \ = \ \pm \ i \ \epsilon \ ,
\eeq
so that there are no solutions of these Killing spinor equations with a due real profile for $\phi$. Similar considerations hold for the NS-NS $h$--cases.

The solutions that we have displayed are supported by an internal flux. They will play a role in the non--supersymmetric backgrounds that we shall explore in Section~\ref{sec:T0hn0}, where they capture limiting behaviors close to a singularity.

\section{\sc Vacuum Solutions and Cosmologies without Form Fluxes} \label{sec:susybnoT}

In this section we construct all the exact vacuum solutions within the class of metrics~\eqref{metric_sym_intro} with $k=k'=0$, an $r$-dependent dilaton profile and no form fluxes. As we just saw, the non--trivial solutions within this class will necessarily break supersymmetry. Then, via suitable analytic continuations, we derive from them corresponding classes of cosmological solutions. 

\subsection{\sc Static Backgrounds}

In this case eqs.~\eqref{EqA_red}, \eqref{EqC_red} and \eqref{Eqphi_red} reduce to
\beq
A'' \ =\  0 \ , \qquad C''\  =\  0 \ , \qquad \phi'' = 0\ ,
\eeq
and therefore the general solution in the harmonic gauge takes the form
\bea
A &=& A_1\,r \,+\, A_2 \ , \qquad B \ = \  (p+1)A+(D-p-2)C \ , \nonumber \\
C &=& C_1\,r \,+\, C_2  \ , \qquad \phi \ = \  \phi_1\,r \ + \ \phi_2 \ ,
\eea
where the $A_i$, $C_i$, $\phi_i$ are arbitrary constants. The constants $A_2$ and $C_2$ can be removed by rescaling all coordinates, thus bringing the solution to the form
\bea{}{}{}{}
ds^2 &=& e^{2 A_1 r}\, dx^2\ + \ e^{2\,\mu\, r}\, dr^2 \ + \ e^{2 C_1 r}\, d\vec{y}^{\,2}  \ , \nonumber \\
e^\phi &=& e^{\phi_1 r + \phi_2} \ , \label{nofluxABC}
\eea
where for simplicity we retain the same symbols, and where
\beq{}{}{}{}
\mu \ = \ (p+1)A_1+(D-p-2)C_1 \ . \label{mu_eq}
\eeq

The Hamiltonian constraint~\eqref{EqB_red} reduces in this case to the homogeneous quadratic form
\beq
\frac{4 \phi_1^2}{D-2} \ = \ p(p+1)A_1^2 \ + 2(p+1)(D-p-2) A_1 C_1 + (D-p-2)(D-p-3) C_1^2 \ . \label{ham_00case}
\eeq
and the special choice $A_1=C_1=\phi_1=0$ corresponds to the supersymmetric flat--space vacuum. Let us also record two useful alternative presentations of this constraint,
\beq
\frac{4 \phi_1^2}{D-2} \ + \ (p+1) \, A_1^2 \ + \ (D-p-2) \, C_1^2 \ = \ \left[ (p+1)A_1 \ + \ (D-p-2)C_1\right]^2 \   \label{ham_00case2}
\eeq
and
\beq
\frac{4\,p\,\phi_1^2}{(p+1)(D-2)} \ = \ \bigg[p\,A_1 \ + (D-p-2)C_1\bigg]^2 \ - \ \frac{(D-2)(D-p-2)}{p+1}\, C_1^2 \ . \label{ham_00case_3}
\eeq

Eq.~\eqref{ham_00case2} shows that, away from the flat--space solution, the values of $A_1$ and $C_1$ are subject to the condition that
$\mu \neq 0$. Therefore, letting
\beq
\alpha_A \ = \ \frac{A_1}{\mu} \ ,  \qquad
\alpha_C \ = \ \frac{C_1}{\mu} \ , \qquad
\alpha_\phi \ = \ \frac{\phi_1}{\mu} \ ,
\eeq
the $\alpha_i$ are determined by the original Hamiltonian constraint~\eqref{ham_00case2} divided by $\mu^2$,
\beq
(p+1) \alpha_A^2 \ + \ (D-p-2) \alpha_C^2 \ + \ \frac{4\, \alpha_\phi^2}{D-2} \ = \ 1 \ , \label{quad_constr}
\eeq
which defines an ellipsoid, while the definition of $\mu$ turns into
\beq
(p+1) \alpha_A \ + \ (D-p-2) \alpha_C \ = \ 1 \ , \label{linear_constr}
\eeq
which describes a plane.
The independent geometries in this class are thus determined by their intersections, which are the points of the ellipse
\beq{}{}{}{}
\frac{(p+1)(D-1)^2}{(D-2)(D-p-2)} \left(\alpha_A \ - \ \frac{1}{D-1}\right)^2 \ + \ \frac{4 (D-1)}{(D-2)^2}\, \alpha_\phi^2 \ =  \ 1 \ . \label{ellipse}
\eeq

We have thus obtained a family of solutions
\bea
ds^2 &=& e^{2\,\alpha_A\,\mu\,r} \, dx^2 \ + \  e^{2\,\mu\,r}\,d r^2\ + \ e^{2\,\alpha_C\,\mu\,r} \, d\vec{y}^{\,2} \ , \nonumber \\
e^\phi &=& e^{\alpha_\phi\,\mu\,r} \, e^{\phi_2} \ , \label{spontaneous_r}
\eea
where $\mu$ is an arbitrary nonzero constant and 
\bea{}{}{}{}
\alpha_A &=& \frac{1}{D-1}\left[1 \ + \ \sqrt{\frac{(D-2)(D-p-2)}{(p+1)}} \ \cos\theta\right] \ , \nonumber \\
\alpha_C &=& \frac{1}{D-1}\left[1 \ - \ \sqrt{\frac{(D-2)(p+1)}{(D-p-2)}} \ \cos\theta\right] \ , \nonumber \\
\alpha_\phi &=& \frac{D-2}{2 \sqrt{D-1}}\ \sin\theta \ \equiv \ \frac{2}{\gamma_c} \ \sin \theta \ , \label{param_theta}
\eea
which are thus parametrized by an angle $\theta$, by the constant $\phi_2$ that enters the dilaton profile and by the scale $\mu$, and where
\beq
\gamma_c \ = \ \frac{4\,\sqrt{D-1}}{D-2} \ , \label{gammac}
\eeq
equals $\frac{3}{2}$ in ten dimensions. For any given choice of $\alpha_A$ and $\alpha_C$, there are pairs of solutions that are mapped into one another by $\theta \to - \theta$, which turns $\alpha_\phi$ to $- \alpha_\phi$. This is the $S$--duality transformation $g_s \to \frac{1}{g_s}$, which links the type-I and $SO(32)$  heterotic strings and maps the type-IIB string into itself.  Particular cases are the two isotropic nine--dimensional solutions, with $\alpha_A=\alpha_C$, which are obtained for $\theta=\pm \frac{\pi}{2}$, for which
\bea
ds^2 &=& e^{\frac{2\,\mu\,r}{D-1} }\, \left( dx^2 \ + \  d\vec{y}^{\,2}\right) \ + \ e^{2\,\mu\,r}\,d r^2 \ , \nonumber \\
e^\phi &=& e^{\,\pm\,\frac{2\,\mu\,r}{\gamma_c}} \, e^{\phi_2} \ , \label{spontaneous_r_isotropic}
\eea
and the constant dilaton solutions, for which $\theta = 0,\pi$ and
\bea{}{}{}{}
\alpha_A &=& \frac{1}{D-1}\left[1 \ \pm \ \sqrt{\frac{(D-2)(D-p-2)}{(p+1)}} \right] \ , \nonumber \\
\alpha_C &=& \frac{1}{D-1}\left[1 \ \mp \ \sqrt{\frac{(D-2)(p+1)}{(D-p-2)}} \right] \ , 
\label{nodilaton}
\eea
which are conventional Kasner solutions. Note that for $D=11$ these expressions provide supersymmetry breaking solutions to the eleven--dimensional supergravity. On the other hand, the $D=2$ case is special, since the solutions reduce to Rindler space.

The backgrounds depend apparently on the three quantities $\mu$, $\theta$ and $\phi_2$. Note, however, that the redefinition
\beq{}{}{}{}
r \ = \ \lambda\left(\widetilde{r} \ + \ r_0 \right) \ , 
\eeq
where $\lambda$ is an arbitrary positive constant and $r_0$ is determined by
\beq
e^{2\mu \lambda r_0}\, \lambda^2 \ = \ 1 \ ,
\eeq
turns the background into
\bea
ds^2 &=& e^{2\,\alpha_A\,\widetilde{\mu}\,\widetilde{r}} \, d\widetilde{x}^2 \ + \  e^{2\,\,\widetilde{\mu}\,\widetilde{r}}\,d \widetilde{r}^2\ + \ e^{2\,\alpha_C\,\widetilde{\mu}\,\widetilde{r}} \, d\vec{\widetilde{y}}^{\,2} \ , \nonumber \\
e^\phi &=& e^{\alpha_\phi\,\widetilde{\mu}\,\widetilde{r}} \, e^{\widetilde{\phi}_2}  \label{spontaneous_rtilde} \ .
\eea
These expressions are of the same form, up to rescalings of the $x$ and $y$ coordinates, but with different values of the parameters, since now
\bea{}{}{}{}
\widetilde{\mu} &=& \lambda\, \mu \ , \nonumber \\
e^{\widetilde{\phi}_2} &=& e^{\phi_2}\, \lambda^{\,-\,\alpha_\phi} \ = \ e^{\phi_2}\, \lambda^ {\,-\,\frac{2}{\gamma_c}\,\sin\theta} \ . \label{identifications}
\eea
Therefore, the triplets $\Big(\mu,\theta,\phi_2 \Big)$ and $\Big(\widetilde{\mu},{\theta},\widetilde{\phi_2} \Big)$ are equivalent for arbitrary nonzero values of $\lambda$. Note also that the backgrounds in eqs.~\eqref{spontaneous_r} include flat space, which corresponds to $\mu = 0$ and is supersymmetric. As we have just seen, they also include another branch, which is not supersymmetric and is parametrized  by the triplets $\Big(\mu,\theta,\phi_2\Big)$, with $\mu \neq 0$, up to the identifications \eqref{identifications}. 

For positive values of $\mu$ a singularity is always present, at a finite distance $\frac{1}{\mu}$ from the origin, as $r \to - \infty$. In this case, one can also define the new variable
\beq
\xi \ = \ \frac{1}{\mu}\ e^{\mu\,r} \ , \label{rxi}
\eeq
where clearly $0 < \xi < \infty$, which turns the background into the space--dependent Kasner--like form
\bea
ds^2 &=& \left(\mu\xi\right)^{2\,\alpha_A} \, dx^2 \ + \ d \xi^2\ + \ \left(\mu \xi\right)^{2\,\alpha_C} \, d\vec{y}^{\,2} \ , \label{spontaneous} \nonumber \\
e^\phi &=& \left(\mu\xi\right)^{\alpha_\phi} \, e^{\phi_2} \ ,
\eea
where the singularity now lies at $\xi=0$. For nonzero values of $\mu$, rescaling the $x$ and $y$ coordinates and redefining $\phi_2$ removes $\mu$ completely. However, this new coordinate system was reached by the transformation~\eqref{rxi}, which becomes singular as $\mu \to 0$, so that the supersymmetric solution is lost in this way. On the other end, the presentation~\eqref{spontaneous_r} has the virtue of encompassing both the non--supersymmetric solutions and the supersymmetric one. In this sense, $\mu$ can be regarded as determining the scale of supersymmetry breaking. The case of eleven--dimensional supergravity is captured again by solutions with $\alpha_{\phi}=0$.

Finally, let us stress that the quantity $\gamma_c$ in eq.~\eqref{gammac} will play an important role in connection with dilaton tadpole potentials. It signals the transition to the climbing behavior~\cite{dks} of corresponding cosmologies but, as we shall see in~\cite{ms_vacuum_2}, it also affects in an important fashion the general dynamics of these systems. 
\subsection{\sc Cosmological Backgrounds}

The analytic continuation $r \to i\,\tau$, and corresponding redefinitions $A_1 \to - i\,A_1$, $C_1 \to - i\,C_1$  and $ \phi_1 \to - i\,\phi_1$, after the transition to the cosmic time $t$, yield 
\bea
ds^2 &=& -
  dt^2
   +
    t^{2\,\alpha_A} \, dx^2 \ + \ t^{2\,\alpha_C} \, d\vec{y}^{\,2} \ , \nonumber \\
  e^\phi &=& t^{\alpha_\phi} \, e^{\phi_2}  \ ,
\eea
where the parameters are still determined by eqs.~\eqref{param_theta}. These are effectively Kasner--like solutions arising from $D+1$ dimensions where, if $\alpha_\phi\neq 0$, the dilaton spans the whole real axis during the cosmological evolution, moving toward larger or smaller values depending on the free sign choice for $\alpha_\phi$ in eq.~\eqref{quad_constr}. There is an interesting option, just like in a conventional Kasner Universe, which is allowed by eq.~\eqref{linear_constr} for $\frac{1}{p+1} < \alpha_A < \frac{1}{\sqrt{p+1}}$: an expansion in the $x$--directions can be accompanied by a contraction in the $y$--directions.

\section{\sc Vacuum Solutions with Form Fluxes} \label{sec:T0hn0}

In Section~\ref{sec:killing_spinors} we saw that form fluxes allow an interesting class of supersymmetric vacua. We can now extend the discussion, exploring the general class of solutions that are allowed by eqs.~\eqref{EqA_red}--\eqref{EqB_red}, and in two particular cases by eqs.~\eqref{eqABC_sdual} and \eqref{ham_sdual}, or by eqs.~\eqref{eqABC_sdual_h} and \eqref{ham_sdual_h}, in the presence of form fluxes of the $H$ and $h$ types, in the nomenclature of Section~\ref{sec:tensor_profiles}. 

Before exploring these solutions, let us recall again that in ten dimensions the $H$--contributions can be associated to RR forms in the type--IIB string when $p$ is odd, and to RR forms in the type--IIA string when $p$ is even, and the $p=1,5$ cases are also relevant for the $SO(32)$ type-I~\cite{greenschwarz} orientifold~\cite{orientifolds} model. Moreover, in $D=10$ these systems can also encompass the fluxes associated to the fundamental string and to the NS fivebrane, in the type--II and heterotic strings, where they can be present. Finally, one can adapt these considerations even to the eleven--dimensional supergravity~\cite{cjs}, where $p=2,5$, setting $\beta_p=0$ and confining the attention to solutions with a vanishing dilaton. We first discuss the $H$--fluxes, to then return at the end of the section to the $h$--fluxes. 

\subsection{\sc Solutions with $H$--Fluxes}~\label{h_flux_sols}

Let us begin again by spelling out the basic ingredients for arbitrary values of $D$, before specializing to the most relevant options.
For this class of solutions, it is convenient to work in terms of the three variables $(C,\phi,Z)$, where
\beq
Z \ = \ (p+1) A \ + \ \beta_p\, \phi \ ,
\eeq
is related to the exponent that accompanies the flux contributions. The relevant values of $\beta_p$ were given in Section~\ref{sec:symmetries}. This choice leads to the three equations
\bea
C''& =&  - \  \frac{(p+1)}{2\,(D-2)} \ e^{\,2\,Z} \ H_{p+2}^2\ , \nonumber \\
\phi'' &=& \frac{\beta_p\,(D-2)}{8}\ e^{\,2\,Z} \ H_{p+2}^2\ , \nonumber \\
Z'' &=& \frac{H_{p+2}^2}{2(D-2)}\left[\left(\frac{\beta_p\,(D-2)}{2}\right)^2 \,+\, (D-p-3)(p+1) \right]e^{\,2\,Z} \ . \label{CphiZ_space}
\eea
Notice that there are linear combinations of $C$ and $Z$, and of $\phi$ and $Z$, whose second derivative with respect to $r$ vanishes identically. Consequently, letting
\beq
\delta^2 \ = \ \frac{1}{(D-2)}\left[\left(\frac{\beta_p\,(D-2)}{2}\right)^2 \,+\, (D-p-3)(p+1) \right] \ ,
\eeq
one can relate $C$ and $\phi$ to $Z$ according to
\bea
C &=& - \ \frac{(p+1)}{(D-2)\,\delta^2} \  Z \ + \ C_1\,r \ + \ C_2 \ , \nonumber \\
\phi &=& \frac{\beta_p\, (D-2)}{4\,\delta^2}\  Z \ + \ \phi_1\,r \ + \ \phi_2\ , \label{CphiZh}
\eea
where $C_{1,2}$ and $\phi_{1,2}$ are integration constants, and finally these results determine $A$ and $B$ as
\bea
A &=& \frac{(D-p-3)}{(D-2)\,\delta^2}\  Z \ - \ \frac{\beta_p}{p+1} \left( \phi_1\,r \ + \ \phi_2\right) \ , \label{ABh} \\
B &=& - \ \frac{(p+1)}{(D-2)\,\delta^2}\  Z \, - \, {\beta_p} \left(\phi_1\,r + \phi_2\right) \, + \, (D-p-2)(C_1\,r+C_2) \ . \nonumber
\eea

The Hamiltonian constraint~\eqref{EqB_red} now gives
\beq
(Z')^2 \ = \ \Delta^2 \ e^{\,2\,Z} \ + \ E \ , \label{encons}
\eeq
where
\beq
\Delta^2 \ = \ \frac{H_{p+2}^2\, \delta^2}{2}   \ .
\eeq
and
\beq
E \ = \ \delta^2\,\Bigg[ \left(\frac{p\,\beta_p^2}{p+1} \ - \ \frac{4}{D-2} \right)\phi_1^2 \ + \ (D-p-2)C_1 \left[ (D-p-3)C_1 \ - \ 2\,\beta_p\,\phi_1\right]\Bigg] \ . \label{energy_h}
\eeq
Eq.~\eqref{encons} has the form of an energy--conservation condition for a Newtonian particle moving in one dimension with total energy $E$.
It corresponds to one of the options discussed in Appendix~\ref{app:deq}, with $\epsilon=1$, and thus with ``negative potential energy''. As explained in Appendix~\ref{app:deq}, the value of the ``energy'' $E$ determines the form of $Z$. Letting
\beq
e^{\,-\,Z} \ = \ f(r) \ , \label{Zf}
\eeq
the different cases are all encompassed by
\beq{}{}
f(r) \ = \ \frac{\Delta}{\sqrt{E}}\, \sinh\left(\sqrt{E} r\right) \ , \label{fE}
\eeq
up to a translation of $r$. In detail, the result for $E=0$
\beq{}{}
f(r) \ = \ {\Delta}\,r \ ,
\eeq
can be recovered as a limit, while the result for $E<0$, 
\beq{}{}
f(r) \ = \ \frac{\Delta}{\sqrt{\left|E\right|}}\, \sin\left(\sqrt{\left|E\right|} r\right) \ .
\eeq
can be recovered by an analytic continuation.
In all cases, one should confine the range of $r$ to a region where $f > 0$. This is the half-line $r > 0$ in the first two cases and the interval $ 0 < r < \frac{\pi}{\sqrt{\left|E \right|}}$ in the last case.

Using eqs.~\eqref{CphiZh}, \eqref{ABh} and \eqref{Zf}, the solutions for the different fields take the form
\bea
ds^2 &=& \left[f(r)\right]^{\,-\,2\,\sigma_A} \, e^{\,-\,\frac{2\,\beta_p\left(\phi_1\,r\,+\,\phi_2\right)}{p+1}}dx^2 \ + \ \left[f(r)\right]^{\,2\,\sigma_B}  \, e^{\,-\,{2\,\beta_p (\phi_1\,r+\phi_2)}+2(D-p-2)(C_1\,r+C_2)}dr^2 \nonumber \\ &+& \left[f(r)\right]^{\,2\,\sigma_B} \, e^{\,2(C_1\,r+C_2)}\,d\vec{y}^{\,2} \ , \nonumber \\
e^\phi &=& \left[f(r)\right]^{\,-\,\sigma_\phi}\, e^{\, \phi_1\,r+\phi_2}\ , \nonumber \\
{\cal H}_{p+2} &=& H_{p+2}\, \frac{\epsilon_{p+1}\, dr}{\left[f(r)\right]^2} \ , \label{gen_sol_hnoTg}
\eea
where
\beq
\sigma_A \,=\, \frac{(D-p-3)}{\delta^2\,(D-2)} \ , \qquad
\sigma_B \,=\,  \frac{(p+1)}{\delta^2\,(D-2)}\ , \qquad \sigma_\phi \,=\, \frac{\beta_p\,(D-2)}{4\,\delta^2} \ . \label{sigmasD}
\eeq
Note that the values of $\beta_p$ given in Section~\ref{sec:symmetries}, with the integration constants and the energy $E$ related as in eq.~\eqref{energy_h}, imply that $\sigma_\phi$ can have both signs in the cases of interest. We can now discuss, in sequence, the cases of $D=11$ and $D=10$. In the latter case we treat separately the self--dual $p=3$ solutions.

\subsubsection{\sc Solutions of Eleven--Dimensional Supergravity}

These solutions are simpler, and can be obtained from eqs.~\eqref{gen_sol_hnoTg} setting $D=11$ and $\beta_p=0$, while also removing the dilaton, so that $\phi_1=0$. Note that $\delta^2=2$ for the two relevant cases $p=2,5$, and now the energy $E$ in eq.~\eqref{energy_h} becomes
\beq{}{}{}{}
E \ = \ 2 (8-p)(9-p) C_1^2 \ ,
\eeq
and is thus always positive. Consequently, one finds
\beq{}{}
f(r) \ = \ \frac{\left|H_{p+2}\right|}{\sqrt{2(9-p)(8-p)}\left|C_1\right|}\, \sinh\left(\sqrt{2(9-p)(8-p)}\left|C_1\right| \, r\right) \  \label{fE11} ,
\eeq
while
\bea
ds^2 &=& \left[f(r)\right]^{\,-\,\frac{8-p}{9}} \,dx^2 \ + \ \left[f(r)\right]^{\,\frac{p+1}{9}}  \left[ e^{\,2(9-p)(C_1\,r+C_2)}dr^2 \nonumber \,+\, e^{\,2(C_1\,r+C_2)}\,d\vec{y}^{\,2} \right] \ , \nonumber \\
{\cal H}_{p+2} &=& H_{p+2}\, \frac{\epsilon_{p+1}\, dr}{\left[f(r)\right]^2} \ . \label{gen_sol_hnoTg11_2}
\eea
In the limit $C_1 \to 0$
\beq{}{}
f(r) \ = \ {\left|H_{p+2}\right|\,r} \  \label{fE112} .
\eeq
and one recovers the supersymmetric solutions described in Section~\ref{sec:killing_spinors}, eqs.~\eqref{solp2} and \eqref{solp5}.

These solutions are to be considered for $r>0$, and there are singularities at $r=0, \infty$, which are separated by a finite distance for $C_1<0$ and by an infinite distance if $C_1 \geq 0$. In both cases of interest, with $p=2,5$, the integral of $\star\,{\cal H}_{p+2}$ over the internal torus is $H_{p+2}$ times the ``parametric'' volume of the torus, proportional to $R^{9-p}$, where $R$ denotes the range of the $y$-coordinates. This result does not depend on $r$, as demanded by the equation for the form, but is not zero and its limiting value as $r\to 0$, where the internal torus associated to the $y$-coordinate shrinks to a point, reveals the presence of a charged extended object with a $(p+1)$-dimensional world volume. In Section~\ref{sec:probe_brane} we shall gather further indications on this type of extended objects, resorting to a probe brane.

As $r\to 0$, the general solution in eqs.~\eqref{gen_sol_hnoTg11} approaches the supersymmetric results that were obtained in Section~\ref{sec:killing_spinors}. On the other hand, as $r \to \infty$ for $C_1 \neq 0$, letting
\beq{}{}{}{}
\sigma \ = \ \sqrt{2(9-p)(8-p)} \ ,
\eeq
\bea
ds^2 &\sim& e^{\,-\,\frac{8-p}{9}\,\sigma\left|C_1\right| r} \,dx^2 \ + \ e^{\,\frac{p+1}{9}\,\sigma\left|C_1\right| r}  \left[ e^{\,2(9-p)\,C_1\,r}dr^2 \nonumber \,+\, e^{\,2\,C_1\,r}\,d\vec{y}^{\,2} \right] \ , \nonumber \\
{\cal H}_{p+2} &\sim & H_{p+2}\, {\epsilon_{p+1}\, dr}\, e^{\,-\,2\,\sigma\left|C_1\right|r} \ . \label{gen_sol_hnoTg11}
\eea
Therefore, the form flux disappears, and the comparison with eqs.~\eqref{spontaneous_r} shows that
the solutions approach the Kasner--like backgrounds that we discussed in the previous section,
with exponents 
\bea
\alpha_A &=& - \, \frac{\left(8-p\right)\sigma}{\left(p+1\right)\sigma \ + \ {18\,\epsilon}\left(9-p\right)} \ , \nonumber \\
\alpha_C &=&  \frac{\left(p+1\right)\sigma \ + \ 18\,\epsilon}{\left(p+1\right)\sigma \ + \ {18\,\epsilon}\left(9-p\right)} \ ,
\eea
where $\epsilon$ is the sign of $C_1$.
In particular, for $p=2$
\beq{}{}{}{}
\alpha_A \ = \  \frac{1}{10}\left(1 \ - \ \epsilon\,\sqrt{21}\right)\ , \qquad \alpha_C \ = \ \frac{1}{10}\left( 1\ + \  \frac{9\,\epsilon}{\sqrt{21}} \right) \ ,
\eeq
while for $p=5$
\beq{}{}{}{}
\alpha_A \ = \  \frac{1}{10}\left(1 \ - \ \epsilon\,\sqrt{6}\right) \ , \qquad \alpha_C \ = \ \frac{1}{10}\left( 1\ + \  \frac{3}{2}\,\epsilon\,\sqrt{6} \right) \ ,
\eeq
and agree with the values given by eqs.~\eqref{nodilaton} with $D=11$ and $p=2,5$. Here we are working in the region $r>0$, so that $C_1<0$ corresponds to a finite length  of the internal interval, while $C_1>0$ corresponds to an infinite length of the internal interval. 

The basic lesson of this setup is clearly visible from these examples: the flux introduces a singularity at $r=0$, where the background approaches a supersymmetric configuration, while also halving the range of $r$. As $r \to + \infty$, for $C_1\neq 0$ these backgrounds approach the no--flux Kasner--like behavior of Section~\ref{sec:susybnoT}, but the presence of fluxes makes it possible to attain a finite length for the internal interval.

\subsubsection{\sc Solutions for $p=3$ and $D=10$, the Self--Dual Case for Type IIB}

This special case is characterized by $\beta_p=0$, and the starting point is provided by eqs.~\eqref{eqABC_sdual}, which imply
\beq{}{}{}
\phi \,=\, \phi_1 \,r \ + \ \phi_2 \ , \qquad
C \,=\, - \ A \ - \ \alpha \, r \ - \ \beta \ ,
\eeq
where $\phi_1$, $\phi_2$, $\alpha$ and $\beta$ are four constants.
Letting now $Y = e^{-4A}$, the Hamiltonian constraint~\eqref{ham_sdual} becomes
\beq{}{}{}
\left( Y'\right)^2 \ - \ 40 \left(\alpha^2 \,-\, \frac{\phi_1^2}{40}\right) Y^2 \ = \ \frac{1}{2}\, H_5^2 \ ,
\eeq
and there are three classes of solutions, which can be studied following Appendix~\ref{app:deq}.
\begin{itemize}
    \item[1. ] If $\alpha^2 \ - \ \frac{\phi_1^2}{40} \ = \ \widetilde{\alpha}^2 \ > \ 0$, the Hamiltonian constraint describes an inverted harmonic oscillator with positive total energy, and the solution reads
    \beq{}{}{}
    Y \ = \ \frac{{H_5}\,\rho}{\sqrt{2}}\, \sinh\left(\frac{r}{\rho}\right) \ ,
    \eeq
    where $0 < r < \infty$, and
    \beq{}{}{}
    \rho \ = \ \frac{1}{2\,\widetilde{\alpha} \, \sqrt{10}} \ .
    \eeq
    Consequently
    \bea{}{}{}
    ds^2 &=& \left(Y\right)^{\,-\,\frac{1}{2}}\, dx^2 \, +\,\left(Y\right)^{\,\frac{1}{2}}\left[ e^{\,-\,5\,\epsilon\,\frac{ \sqrt{1\,+\,\left(\phi_1\,\rho\right)^2}}{\sqrt{10}}\,\frac{r}{\rho} \,-\,10\,\beta} \, dr^2\ + \ e^{\,-\,\epsilon \,\frac{\sqrt{1\,+\,\left(\phi_1\,\rho\right)^2}}{\sqrt{10}}\,\frac{r}{\rho}}\,d\vec{y}^{\,2} \right] \ , \nonumber \\
  {\cal H}_{5} &=& \frac{H_{5}}{2 \sqrt{2}}\,\left(\frac{{\epsilon}_{(4)}  \, dr}{Y^2}  \,+\, e^{5\,\beta}\,\widetilde{\epsilon}_{(5)} \right)   \ , \qquad
  e^\phi \,=\, e^{\phi_1\,r \,+\, \phi_2} \ . \label{fiveformEpos}
    \eea
    where $\epsilon= \pm 1$ is the sign of $\alpha$. These solutions depend on $\phi_1$, $\phi_2$, $\beta$ and involve finite values of $\rho$. However, rescalings of the $x$ and $y$ coordinates and redefinitions of $\phi_1$ so that $\phi_1\rho = \widetilde{\phi}_1$, and of $\beta$ so that $\beta=\widetilde{\beta}+\frac{1}{4}\,\log\rho$, can eliminate the dependence on $\rho$ altogether, thus leaving a three--parameter family of solutions.  Negative {or zero} values of $\phi_1$ grant a bounded string coupling, while positive values of $\alpha$ grant a finite internal length.
    \item[2. ] If $\alpha^2 \ - \ \frac{\phi_1^2}{40} \ = \ 0$, the solution is simply
     \beq{}{}{}
    Y \ = \ \frac{{H_5}\,r}{\sqrt{2}} \ ,
    \eeq
    where $0 < r < \infty$, and consequently
    \bea{}{}{}
    ds^2 &=& \left(Y\right)^{\,-\,\frac{1}{2}}\, dx^2 \, +\,\left(Y\right)^{\,\frac{1}{2}}\left[ e^{\,-\,\frac{5 \epsilon \left|\phi_1\right|r}{\sqrt{10}}\,-\,10\,\beta} \, dr^2\ + \ e^{\,-\,\frac{\epsilon \left|\phi_1\right|r}{\sqrt{10}}}\,d\vec{y}^{\,2} \right] \ , \nonumber \\
  {\cal H}_{5} &=& \frac{H_{5}}{2 \sqrt{2}}\,\left(\frac{{\epsilon}_{(4)}  \, dr}{Y^2}  \,+\, e^{5\,\beta}\,\widetilde{\epsilon}_{(5)} \right)   \ , \qquad
  e^\phi \,=\, e^{\phi_1\,r \,+\, \phi_2} \ .
    \eea
    These expressions are limiting forms of eqs.~\eqref{fiveformEpos} as $\rho \to \infty$, and depend on the three parameters $\phi_1$, $\phi_2$ and $\beta$. For $\phi_1=0$, they reduce to the supersymmetric solutions of Section~\ref{sec:killing_spinors}, and in this case $\beta$ can be eliminated rescaling the coordinates. Negative values of $\phi_1$ grant a bounded string coupling, while positive values of $\alpha$ grant again a finite internal length.
    
 \item[3. ] If $\alpha^2 \ - \ \frac{\phi_1^2}{40} \ = \ - \ \widetilde{\alpha}^2 \ < \ 0$, the Hamiltonian constraint describes an ordinary harmonic oscillator, and the solution reads
     \beq{}{}{}
    Y \ = \ \frac{{H_5}\,\rho}{\sqrt{2}}\, \sin\left(\frac{r}{\rho}\right) \ ,
    \eeq
    where $0 < r < \pi\,\rho$, with
    \beq{}{}{}
    \rho \ = \ \frac{1}{2\,\widetilde{\alpha} \, \sqrt{10}} \ .
    \eeq
    Now $\phi_1^2\, \rho^2 > 1$, and the solution reads
        \bea{}{}{}
    ds^2 &=& \left(Y\right)^{\,-\,\frac{1}{2}}\, dx^2 \, +\,\left(Y\right)^{\,\frac{1}{2}}\left[ e^{\,-\,\frac{5 \epsilon \sqrt{\left(\phi_1\,\rho\right)^2\,-\,1}}{\sqrt{10}}\,\frac{r}{\rho} \,-\,10\,\beta} \, dr^2\ + \ e^{\,-\,\frac{\epsilon \sqrt{\left(\phi_1\,\rho\right)^2}\,-\,1}{\sqrt{10}}\,\frac{r}{\rho}}\,d\vec{y}^{\,2} \right] \ , \nonumber \\
  {\cal H}_{5} &=& \frac{H_{5}}{2 \sqrt{2}}\,\left(\frac{{\epsilon}_{(4)}  \, dr}{Y^2}  \,+\, e^{5\,\beta}\,\widetilde{\epsilon}_{(5)} \right)   \ , \qquad
  e^\phi \,=\, e^{\phi_1\,r \,+\, \phi_2} \ .
    \eea
    As in the first case, one can eliminate $\rho$, and this is therefore a three--parameter family of solutions, with a bounded string coupling for all values of $\phi_1$ since the range of $r$ is bounded. The length of the internal interval is always bounded in this case.
\end{itemize}

As $r \to 0$ the three families of solutions approach the supersymmetric ones of Section~\ref{sec:killing_spinors}. This is also true at the other end for the third type of solutions. As $r\to\infty$ the solutions of the first type approach the Kasner--like behavior of Section~\ref{sec:susybnoT}, with
\bea{}{}{}{}
\alpha_A &=& - \ \frac{1}{1 \ - \ \epsilon\,\sqrt{10}\,\sqrt{1 \,+\, 
\widetilde{\phi}_1^{\,2}}}\ , \nonumber \\
\alpha_C &=& \frac{1 \ - \ \frac{\epsilon}{5}\,\sqrt{10}\,\sqrt{1 \,+\, \widetilde{\phi}_1^{\,2}}}{1 \ - \ {\epsilon}\,\sqrt{10}\,\sqrt{1 \,+\, \widetilde{\phi}_1^{\,2}}} \ , \nonumber \\
\alpha_\phi &=&  \frac{4\,\widetilde{\phi}_1}{1 \ - \ \epsilon\,\sqrt{10}\,\sqrt{1 \,+\, \widetilde{\phi}_1^{\,2}}} \ ,
\eea
which satisfy the constraints of eqs.~\eqref{quad_constr} and \eqref{linear_constr}. The corresponding relations for the second case with $\phi_1 \neq 0$ can be obtained from these expressions in the limit $\widetilde{\phi}_1 \to \infty$, so that the behavior as $r \to \infty$ is again Kasner-like, with $(\alpha_A,\alpha_C,\alpha_\phi)=\left(0,\frac{1}{5},\,-\,\epsilon\,\sqrt{\frac{2}{5}}\right)$.

These families of solutions have the very interesting property of allowing, within a certain range of values for the parameters, compactifications to four--dimensional Minkowski space that combine a five--torus with an internal interval of finite length, where the string coupling is everywhere bounded. We shall return to them in~\cite{int4d_vacuum}.

\subsubsection{\sc Solutions for $D=10$ and $p \neq 3, 8$}

In order to discuss the remaining ten--dimensional solutions, we now return to eqs.~\eqref{gen_sol_hnoTg} and \eqref{sigmasD}, which reduce in $D=10$ to
\bea
ds^2 &=& \left[f(r)\right]^{\,\frac{p-7}{8}} \, e^{\,-\,\frac{2\,\beta_p\left(\phi_1\,r\,+\,\phi_2\right)}{p+1}}dx^2 \ + \ \left[f(r)\right]^{\,\frac{p+1}{8}}  \, e^{\,-\,{2\,\beta_p (\phi_1\,r+\phi_2)}+2(8-p)(C_1\,r+C_2)}dr^2 \nonumber \\ &+& \left[f(r)\right]^{\,\frac{p+1}{8}} \, e^{\,2(C_1\,r+C_2)}\,d\vec{y}^{\,2} \ , \nonumber \\
e^\phi &=& \left[f(r)\right]^{\,-\,\beta_p}\, e^{\, \phi_1\,r+\phi_2}\ , \nonumber \\
{\cal H}_{p+2} &=& H_{p+2}\, \frac{\epsilon_{p+1}\, dr}{\left[f(r)\right]^2} \ . \label{gen_sol_hnoT}
\eea
Moreover, in the string frame and for Ramond--Ramond forms this class of solutions reads
\bea
ds_s^2 &=& \left[f(r)\right]^{\,-\,\frac{1}{2}} \, e^{\,\frac{2(\phi_1\,r +\phi_2)}{p+1}}\ dx^2 \ + \ \left[f(r)\right]^{\,\frac{1}{2}}  \, e^{\,\frac{(4-p)(\phi_1\,r+\phi_2)}{2}\,+\,2(8-p)(C_1\,r+C_2)}dr^2 \nonumber \\ &+& \left[f(r)\right]^{\,\frac{1}{2}} \, e^{\,2(C_1\,r+C_2) \,+\, \frac{\phi_1 r + \phi_2}{2}}\,d\vec{y}^{\,2} \ , \nonumber \\
e^\phi &=& \left[f(r)\right]^{\,\frac{3-p}{4}}\, e^{\, \phi_1\,r+\phi_2}\ , \nonumber \\
{\cal H}_{p+2} &=& H_{p+2}\, \frac{\epsilon_{p+1}\, dr}{\left[f(r)\right]^2} 
\label{t0hn0_gen_string}
\eea
while for NS--NS forms
\bea
ds_s^2 &=& \left[f(r)\right]^{\,\frac{p-5}{4}} \, e^{\,\left(\phi_1\,r\,+\,\phi_2\right)\frac{(p-1)}{(p+1)}}\ dx^2 \ + \ \left[f(r)\right]^{\,\frac{p-1}{4}}  e^{\,\frac{(p-2)}{2}\,(\phi_1\,r+\phi_2)+2(8-p)(C_1\,r+C_2)}dr^2 \nonumber \\ &+& \left[f(r)\right]^{\,\frac{p-1}{4}} \, e^{\,2(C_1\,r+C_2) \,+\, \frac{\phi_1 r + \phi_2}{2}}\,d\vec{y}^{\,2} \ , \nonumber \\
e^\phi &=& \left[f(r)\right]^{\,\frac{p-3}{4}}\, e^{\, \phi_1\,r+\phi_2}\ , \nonumber \\
{\cal H}_{p+2} &=& H_{p+2}\, \frac{\epsilon_{p+1}\, dr}{\left[f(r)\right]^2} \ ,
\label{t0hn0_gen2}
\eea
with $p=1,5$. In all these cases $D=10$, and $\delta^2=2$, so that our starting point is
\beq
E \ = \ 2\,\Bigg[ \left(\frac{p\,\beta_p^2}{p+1} \ - \ \frac{1}{2} \right)\phi_1^2 \ + \ (8-p)C_1 \left[ (7-p)C_1 \ - \ 2\,\beta_p\,\phi_1\right]\Bigg] \ . \label{energy_h10}
\eeq
\
For $p \neq 7$, this expression can be cast in the convenient form
\beq
E \ = \ 2\,\Bigg[(8-p)(7-p)\left( C_1 - \frac{\beta_p\,\phi_1}{7-p} \right)^2  \ - \ \frac{8 \,\phi_1^2}{(p+1)(7-p)} \Bigg] \ , \label{E_value}
\eeq
while if $p=7$
\beq
E \ = \ \frac{3}{4} \left[ \left(\phi_1 \ - \ \frac{8}{3}\, C_1 \right)^2 \ - \ \frac{64}{9}\, C_1^2 \right] \ . \label{E_value7}
\eeq
In order to discuss these solutions further, one must now distinguish three sub--cases, depending on the value of $E$:
\begin{itemize}
\item[1. ] If $ E > 0$ in eq.~\eqref{E_value}, letting $E = \frac{1}{\rho^2}$, one can see from eq.~\eqref{fE} that
\beq
f(r) \ = \ \left|H_{p+2}\right|\,\rho\,\sinh\left(\frac{r}{\rho}\right) \ , \label{fposE}
\eeq
where $0 < r < \infty$, up to a translation and a choice of orientation on the $r$ axis. Eq.~\eqref{E_value} then shows that, for $p<7$, the independent choices for $C_1$ and $\phi_1$ can be expressed in terms of a real parameter $\zeta$ according to
    \bea
    C_1 &=& \frac{\beta_p}{4\,\rho} \, \sqrt{\frac{p+1}{7-p}}\, \sinh \zeta \ \pm \ \frac{\cosh \zeta}{\rho\,\sqrt{2(8-p)(7-p)}} \ , \nonumber \\
    \phi_1 &=& \frac{\sqrt{(p+1)(7-p)}}{4\,\rho}\, \sinh \zeta \ , \label{C1f1Epos}
    \eea
    while for $p=7$
    \beq
    \phi_1 \ = \ \pm \ \frac{2}{\sqrt{3}\,\rho} \ e^{\,\pm\,\zeta} \ , \qquad
    C_1 \ = \ \frac{\sqrt{3}}{4\,\rho}\, \sinh \zeta \ . \label{C1f1Epos7}
    \eeq

\item[2. ] If $E=0$, one can see from Appendix~\ref{app:deq} that
\beq
f(r) \ = \ \left|H_{p+2}\right|\,r \ , \label{fre0}
\eeq
where $0 < r < \infty$, and then for $p<7$
\beq
C_1 \ = \ \frac{\phi_1}{7-p} \left[ \beta_p \ \pm \ \frac{2\,\sqrt{2}\,\sign(\phi_1)}{\sqrt{(p+1)(8-p)}}\right] \ ,
\label{one_param_E0}
\eeq
while for $p=7$ there are again two branches of solutions, with
    \beq
    \phi_1 \ = \ 0 \label{p7E0first}
    \eeq
    and $C_1$ arbitrary, or with
    \beq
    \phi_1 \ = \ \frac{16}{3}\, C_1 \ . \label{p7E0second}
    \eeq
    The solutions with $E=0$ and $\phi_1=0$ are supersymmetric, and were already discussed in Section~\ref{sec:killing_spinors}. 

\item[3. ] If $E<0$, letting $E = \,-\,\frac{1}{\rho^2}$, one can see from Appendix~\ref{app:deq} that
\beq
f(r) \ = \ \left|H_{p+2}\right|\,\rho\,\sin\left(\frac{r}{\rho}\right) \ ,
\eeq
where $0 < r < \pi\,\rho$, up to a translation and a choice of orientation on the $r$ axis. Eq.~\eqref{E_value} then shows that for $p<7$ the independent choices for $C_1$ and $\phi_1$ can be expressed in terms of a real parameter $\zeta$, according to
\bea
\phi_1 &=& \pm \ \frac{\sqrt{(p+1)(7-p)}}{4\,\rho}\, \cosh \zeta \ , \nonumber \\
 C_1 &=& \pm\ \frac{\beta_p}{4\,\rho} \, \sqrt{\frac{p+1}{7-p}}\, \cosh \zeta \ + \ \frac{\sinh \zeta}{\rho\,\sqrt{2(8-p)(7-p)}} \ . \label{paramEneg}
\eea
For $p=7$, there are again two branches of solutions, with
    \beq
    \phi_1 \ = \ \pm \ \frac{2}{\sqrt{3}\,\rho} \ e^{\,\pm\,\zeta} \ , \qquad
    C_1 \ = \ \frac{\sqrt{3}}{4\,\rho}\, \sinh \zeta \ . \label{paramEneg7}
    \eeq
\end{itemize}

Summarizing, this class of backgrounds with fluxes apparently contains five real parameters and a two-valued discrete one, which identifies the two branches. The real parameters are $H_{p+2}$ and $\rho$, both of which enter the function $f$ of eq.~\eqref{fE}, $\zeta$, and the three constants $C_2$, $\phi_2$, leaving aside the moduli of the internal torus.
However, rescaling the coordinates and redefining the constants $C_1$, $C_2$, $\phi_1$ and $\phi_2$ one can remove completely the dependence on both $\rho$ and $H_{p+2}$ in eqs.~\eqref{gen_sol_hnoT}. In conclusion, one is left with $\zeta$, $C_2$ and $\phi_2$, for each of the two branches. These considerations hold for both positive and negative values of $E$, while for $E=0$ one can remove $H_{p+2}$ by a rescaling and then, proceeding as in other case, one ends up again with three parameters, say $\phi_1$, $\phi_2$ and $C_2$.

\subsubsection{\sc Solutions for $D=10$ and $p=8$}

This is a formal limit of the general solutions. In this case $C$ is not present and the energy is positive, with
\beq
E \ = \ \frac{1}{\rho^2} \ = \ \frac{16}{9}\, \phi_1^2 \ ,
\eeq
so that
\beq
f(r) \ = \ \left|H_{10}\right|\,\rho\,\sinh\left(\frac{r}{\rho} \right) \ .
\eeq
Consequently, using
\beq{}{}{}{}{}
\phi_1 \ = \ \pm \ \frac{3}{4\,\rho} \ , 
\eeq
one gets
\bea
ds^2 &=& e^{\,\mp\,\frac{15}{72}\,\frac{r}{\rho}\,-\,\frac{5}{18}\,\phi_2}\left[\left|H_{10}\right|\,\rho\,\sinh\left(\frac{r}{\rho} \right)\right]^\frac{1}{8} dx^2 \ + \ e^{\,\mp\,\frac{15}{8}\,\frac{r}{\rho}\,-\,\frac{5}{2}\,\phi_2}\left[\left|H_{10}\right|\,\rho\,\sinh\left(\frac{r}{\rho} \right)\right]^\frac{9}{8} dr^2\ , \nonumber \\
e^\phi &=& \frac{e^{\,\pm\,\frac{3}{4}\,\frac{r}{\rho}}\ e^{\,\phi_2}}{\left[\left|H_{10}\right|\,\rho\,\sinh\left(\frac{r}{\rho} \right)\right]^\frac{5}{4}} \ , \label{nonsusyD8}\\
{\cal H}_{10} &=& \frac{H_{10}\, \epsilon_{10}}{\left[\left|H_{10}\right|\,\rho\,\sinh\left(\frac{r}{\rho} \right)\right]^2} \ . \nonumber
\eea
In the string frame these results become
\beq
ds_s^2 \,=\, e^{\,\pm\,\frac{1}{6}\,\frac{r}{\rho}\,+\,\frac{2}{9}\,\phi_2}\left[\left|H_{10}\right|\,\rho\,\sinh\left(\frac{r}{\rho} \right)\right]^{\,-\,\frac{1}{2}} dx^2 \ + \ e^{\,\mp\,\frac{3}{2}\,\frac{r}{\rho}\,-\,2\,\phi_2}\left[\left|H_{10}\right|\,\rho\,\sinh\left(\frac{r}{\rho} \right)\right]^\frac{1}{2} dr^2 \ .
\eeq

There are two branches of solutions, in which the string coupling always diverges at the origin and tends to zero as $r \to \infty$. As above, finite values of $\left|H_{10}\right|$ and $\left|\rho\right|$ can be eliminated by redefinitions of $\phi_2$ and rescalings of the coordinates. Hence, this is effectively a one--parameter family with two branches. The length along the $r$ direction and the effective Planck mass are finite for the upper branch and infinite for the lower one. In the former case, the string coupling vanishes at the right end of the interval. There is also a special $E=0$ solution, which can be obtained in the limit $\rho \to \infty$, and reads
\bea
ds^2 &=& \left(\left|H_{10}\right|\,r \right)^\frac{1}{8} dx^2 \ + \ e^{\,-\,\frac{5}{2}\,\phi_2}\,\left(\left|H_{10}\right|\,r\right)^\frac{9}{8} dr^2\ , \nonumber \\
e^\phi &=& \frac{e^{\phi_2}}{\left(\left|H_{10}\right|\,r\right)^\frac{5}{4}} \ , \\
{\cal H}_{10} &=& \frac{H_{10}\, \epsilon_{10}}{\left(\left|H_{10}\right|\,r\right)^2} \ , \nonumber
\eea
while the corresponding string--frame expressions are
\bea
ds_s^2 &=& e^{\,\frac{2}{9}\,\phi_2}\left(\left|H_{10}\right|\,r\right)^{\,-\,\frac{1}{2}} dx^2 \ + \ e^{\,-\,2\,\phi_2}\left(\left|H_{10}\right|\,r\right)^\frac{1}{2} dr^2 \ , \nonumber \\
e^\phi &=& \frac{e^{\,\phi_2}}{\left[\left|H_{10}\right|\,r\right]^\frac{5}{4}} \ .
\eea
This is the familiar supersymmetric $D8$ brane solution of type IIA, which we also encountered in Section~\ref{sec:killing_spinors}, while the backgrounds in eqs.~\eqref{nonsusyD8} are non--supersymmetric deformations of it. Notice also that all these solutions behave in the same fashion near $r=0$, where they approach the supersymmetric background of Section~\ref{sec:killing_spinors}.

\subsubsection{\sc General Properties of the Ten--Dimensional Solutions}

These metrics are always singular at $r=0$. This can be seen from the scalar curvature, which is always singular for $p \neq 3$, or from the invariant $R_{MN} R^{MN}$, which is singular in the latter case. 

\subsubsubsection{\sc Behavior near $r=0$}

Close to $r=0$, the function $f(r)$ of eq.~\eqref{fE} becomes independent of $E$, and
\beq{}{}
f(r) \ \sim \ \left|H_{p+2}\right|\, r \ .
\eeq
The limiting behavior of the solutions is thus, for $p \neq 3$,
\bea
ds^2 &\sim& \left(\left|H_{p+2}\right|\,r\right)^{\,\frac{p-7}{8}} \, dx^2 \ + \ {\left(\left|H_{p+2}\right|\,r\right)^{\,\frac{p+1}{8}}} \left( { e^{\,-\,{2\,\beta_p \phi_2}}\,e^{2(8-p)C_2}\,dr^2 \ + \ d\vec{y}^{\,2}}\right) \ , \nonumber \\
e^\phi &\sim& \left(\left|H_{p+2}\right|\,r\right)^{\,-\,\beta_p} \, e^{\phi_2} \ , \nonumber \\
{\cal H}_{p+2} &\sim& H_{p+2}\, \frac{\epsilon_{p+1}\, dr}{\left(\left|H_{p+2}\right|\,r\right)^2} \ . \label{universal}
\eea
This result is the same for the three classes of backgrounds, while the contributions of $\phi_1$ and $C_1$ are subdominant. Note that this general limiting behavior reproduces the supersymmetric profiles of Section~\ref{sec:killing_spinors}, and indeed eqs.~\eqref{universal} and \eqref{metricsusypE} coincide, up to a rescaling of $r$. This result is what we also found for the eleven--dimensional supergravity and for the self--dual case of type IIB.

The limiting form of the backgrounds close to $r=0$ affords a convenient presentation in terms of a variable $\xi$ that measures the proper length along the $r$--direction. Letting
\beq
\left|H_{p+2}\right|\,\xi \ = \ 16\ e^{\,-\,{\beta_p\,\phi_2}}\,e^{(8-p)C_2}\, \frac{\left(\left|H_{p+2}\right|\,r\right)^{\,\frac{p+17}{16}}}{p+17} \ ,
\eeq
the metric and the string coupling take the simple forms
\bea
ds^2 &=& \left(\left|H_{p+2}\right|\,\xi\right)^{\,-\,\frac{2(7-p)}{(p+17)}} dx \cdot dx  \ + \ d\, \xi^2 \ + \   \left( \left|H_{p+2}\right|\,\xi \right)^{\,{\,\frac{2(p+1)}{(p+17)}} }\, d\,\vec{y}\cdot d\,\vec{y} \ , \nonumber \\
e^\phi &=& \frac{e^{\phi_2}}{\left(\left|H_{p+2}\right|\,\xi\right)^{\frac{16\,\beta_p}{p+17}}} \ , \label{universal_0limit}
\eea
after absorbing some constants in the normalizations of the $x$ and $y$ coordinates, and after a redefinition of $\phi_2$. These expressions define a Kasner--like background but are different from those of Section~\ref{sec:susybnoT}. Notice that the contribution to the length of the interval from the $\xi=0$ end is always finite, so that these spacetimes have a boundary there, where the string coupling can vanish or diverge, depending on the value of $\beta_p$. In particular, for RR forms the string coupling is finite for $p \leq 3$, and the same is true for the NS--NS 7--form field strength.

\subsubsubsection{\sc Behavior for large values of $r$}

On the other hand, the asymptotic behavior for large values of $r$ depends on the energy $E$, so that one must distinguish a number of cases.

\begin{itemize}

\item[1. ] For $E > 0$ and $p<7$, in terms of the proper length $\xi$, as $r \to +\infty$  the metric and string coupling approach
\bea
ds^2 &\sim& \xi^{\,2\,\alpha_A}\, dx^2 \ + \ \, d\xi^2 \ + \ \xi^{\,2\,\alpha_C}\, dy^2 \ , \nonumber \\
e^\phi &\sim& \xi^{\alpha_\phi} \ , \label{kasnerlike}
\eea
with
\beq{}{}{}{}
\alpha_A \ = \ \frac{a}{b} \ , \qquad \alpha_C \ = \ \frac{c}{b} \, \qquad \alpha_\phi \ = \ \frac{f}{b} \ ,
\eeq
where, for $p<7$,
\bea
a &=& - \ \frac{1}{\rho} \left[\frac{7-p}{16} \ + \ \frac{\beta_p}{4} \, \sqrt{\frac{7-p}{p+1}}\, \sinh \zeta \right]  \ , \nonumber \\
b &=&  \frac{1}{\rho} \left[\frac{p+1}{16} \ + \ \frac{\beta_p}{4}\, \sqrt{\frac{p+1}{7-p}}\, \sinh\zeta \ \pm \ \sqrt{\frac{8-p}{2(7-p)}}\, \cosh \zeta\right] \ , \nonumber \\
c &=& \frac{1}{\rho} \left[\frac{p+1}{16} \ + \ \frac{\beta_p}{4}\, \sqrt{\frac{p+1}{7-p}}\, \sinh\zeta \ \pm \ \frac{1}{\sqrt{2(7-p)(8-p)}}\, \cosh \zeta\right]\ , \nonumber \\
f &=&  \frac{1}{\rho} \left[ \ - \ \beta_p \ + \ \frac{1}{4}\ \sqrt{(p+1)(7-p)}\, \sinh\,\zeta\right]  \ . \label{large_r_asympt}
\eea
For $p=7$, with the parametrization~\eqref{C1f1Epos7}, the coefficients become
\bea
a &=& \mp \ \frac{e^{\pm \zeta}}{4\sqrt{3}\,\rho} \ , \nonumber \\
b &=& \frac{1}{\rho} \left[ \frac{1}{2} \ \mp \ \frac{2\ e^{\pm \zeta}}{\sqrt{3}} \ + \ \frac{\sqrt{3}}{4} \ \sinh \zeta \right] \ , \nonumber \\
c &=& \frac{1}{\rho} \left[ \frac{1}{2} \ + \ \frac{\sqrt{3}}{4} \ \sinh \zeta \right] \ , \nonumber \\
f &=& \frac{1}{\rho} \left[  \pm \ \frac{2\ e^{\pm \zeta}}{\sqrt{3}} \ - \ 1 \right] \label{large_r_asympt7} \ .
\eea
$\alpha_A$, $\alpha_C$ and $\alpha_\phi$ satisfy eqs.~\eqref{quad_constr} and \eqref{linear_constr}, so that the large--$r$ behavior of these solutions is captured by the Kasner--like vacua of Section~\ref{sec:susybnoT}. This can be verified from these expressions, but it is implied by the fact that, for $E>0$, $Z \sim \,-\,\frac{r}{\rho}$ as $r \to \infty$. As a result, $A$, $C$ and $\phi$ in eqs.~\eqref{CphiZh} and \eqref{ABh} approach a linear dependence on $r$, while the Hamiltonian constraint~\eqref{energy_h10} is independent of the flux. This is precisely the setting of Section~\ref{sec:susybnoT}.

Depending on the sign of $b$, this type of asymptotic behavior corresponds indeed to one or the other of the two regions $r \to \pm \infty$ of eqs.~\eqref{spontaneous_r}, which describe corresponding solutions in the absence of a flux. On the other hand, close to $r=0$ the flux dominates and spacetime ends. For this reason one can obtain solutions with positive energy, finite string coupling, finite $r$--length and finite effective Planck mass in the presence of fluxes, which was impossible in the setup of Section~\ref{sec:susybnoT}. One can rightfully state that these vacua interpolate between the supersymmetric solutions of Section~\ref{sec:killing_spinors}, in the region where the flux dominates, and the Kasner--like solutions in the absence of fluxes of Section~\ref{sec:susybnoT}, in the region where the flux fades out.

Alternatively, in the absence of flux there was always a boundary, in eq.~\eqref{spontaneous_r}, as $r\to -\infty$, and here the flux introduces another boundary.
If the new boundary coincides with the original one, the resulting setup is similar to the one in Section~\ref{sec:susybnoT}. However, if the new boundary lies at the other end, the result can be a finite interval with the novel features that we have illustrated. 
$B$ determines indeed the length of the interval, which can be finite or infinite, and the same is true for the asymptotic behavior of the string coupling as $r\to+\infty$. Moreover, in the gauge $F=0$ the $(p+1)$--dimensional Planck mass is determined by $B-A$ if the internal $(8-p)$ dimensions are compact, since
\beq
\left(M_{Pl(p+1)}\right)^{p-1} = \ \left(M_{Pl(10)}\right)^{8} \, \mathrm{Vol}(T^{p+1}) \ \times  \ \int_0^\infty dr \ e^{2(B-A)} \ ,
\eeq
where $\mathrm{Vol}(T^{p+1}) = R^{p+1}$ is the parametric volume of the internal torus.
Since asymptotically $B-A \sim (b-a) r$, the finiteness of the $p+1$--dimensional Planck mass is granted by a negative sign of
\beq{}{}{}{}
b\ -\ a \,=\,  \frac{1}{\rho} \left[\frac{1}{2} \ + \ \frac{2\,\beta_p}{\sqrt{(p+1)(7-p)}}\, \sinh \zeta \ \pm \ \sqrt{\frac{(8-p)}{2(7-p)}}\, \cosh \zeta\right]
\eeq
if $p<7$, or by a negative sign of
\beq{}{}{}{}
b \ - \ a  \,=\, \frac{1}{\rho} \left[ \frac{1}{2} \ \mp \ \frac{7\ e^{\pm \zeta}}{4\,\sqrt{3}} \ + \ \frac{\sqrt{3}}{4} \ \sinh \zeta \right]
\eeq
if $p=7$, since the corresponding contributions at the origin are well behaved for all values of $p$. Both options are possible, depending on the branch and the value of $\zeta$, { but for $p>0$ finite(infinite) internal lengths always accompany finite(infinite) values of the effective Planck mass.}

There are also special values of $\zeta$ in both branches, such that $f=0$,
\beq
\sinh \zeta \ = \ \frac{4\,\beta_p}{\sqrt{(p+1)(7-p)}} \ ,
\eeq
for which the string coupling approaches a finite value as $r \to+\infty$. This value separates the two regions of strong and weak coupling for large values of $r$. Hence, \emph{there are cases where supersymmetry is broken, the string coupling is bounded, the length of the $r$--interval is finite and the effective Planck mass are also finite}. 

In particular, for $p=3$, a special case of eqs.~\eqref{fiveformEpos} with $\phi_1=0$ yields a relatively simple four--dimensional flat vacuum of this type from the type--IIB string, where the string coupling is constant, and is thus finite everywhere. In the Einstein frame, with vanishing values for $\phi_2$ and $\beta$, the corresponding metric reads
\bea
ds^2 &=&  \left[\frac{|H_5|}{\sqrt{2}}\,\rho\,\sinh\left(\frac{r}{\rho}\right)\right]^{\,-\,\frac{1}{2}} \!\! dx^2 \nonumber \\ &+&  \left[\frac{|H_5|}{\sqrt{2}}\,\rho\,\sinh\left(\frac{r}{\rho}\right)\right]^{\,\frac{1}{2}}  \left( e^{\,-\,\sqrt{\frac{5}{2}}\,\frac{r}{\rho}} \, dr^2 \, + \, e^{\,-\,\frac{r}{\rho\,\sqrt{10}}} \,d\vec{y}^{\,2}\right)\ \ . \label{4d_inter}
\eea
The corresponding five-form field strength reads
\beq
{\cal H}_5 \ = \ 
\frac{H_5}{2 \sqrt{2}} \left\{ \frac{dx^0 \wedge ...\wedge dr}{\left[\frac{|H_5|}{\sqrt{2}}\,\rho\,\sinh\left(\frac{r}{\rho}\right)\right]^2} \ + \ dy^1 \wedge ... \wedge dy^5\right\} \label{4d_inter_flux} \ .
\eeq
Notice that, in the limit $\rho\to \infty$, these expressions reduce the supersymmetric ones in eqs.~\eqref{killing_data}, which were obtained from the Killing--spinor equations. 

\item[2. ] For $E=0$, $f(r)$ is given in eq.~\eqref{fre0}, and for large values of $r$ and for RR forms, the contributions to the metric behave as
\bea{}{}{}{}
A &\sim& -\ \frac{|\phi_1|\, \epsilon\left(p-3\right) r}{4\left(p+1\right)} \ , \nonumber \\
B & \sim & \frac{\epsilon\,|\phi_1| \,r}{4(7-p)\sqrt{p+1}} \left[(p-3) \sqrt{p+1} \ \pm \ 8\,\epsilon\,\sqrt{2} \,\sqrt{8-p}\right] \ , \nonumber \\
C &\sim& \frac{|\phi_1|\,\epsilon \,r}{4(7-p)\sqrt{p+1}}\left[(p-3)\sqrt{p+1} \ \pm \ \frac{8\sqrt{2}}{\sqrt{8-p}}\right] \ , \nonumber \\
\phi &\sim& \epsilon \,|\phi_1|\, r \ ,
\eea
where $\epsilon$ is the sign of $\phi_1$. 

Consequently, for $\phi_1\neq 0$ the solutions approach, at the second boundary, a no--flux Kasner--like behavior as in eqs.~\eqref{kasnerlike}, with
\bea{}{}{}{}
\alpha_A &=& - \ \frac{(7-p)(p-3)}{\sqrt{p+1}\left[(p-3)\sqrt{p+1} \ \pm \ {8\,\epsilon\,\sqrt{2}}\,\sqrt{8-p}\right]} \ , \nonumber \\
\alpha_C &=&  \frac{(p-3)\sqrt{p+1} \ \pm \ \frac{8\,\epsilon\,\sqrt{2}}{\sqrt{8-p}} }{\left[(p-3)\sqrt{p+1} \ \pm \ {8\,\epsilon\,\sqrt{2}}\,\sqrt{8-p}\right]}  \ , \nonumber \\
\alpha_\phi &=& \frac{4\,\epsilon\,(7-p)\sqrt{p+1}}{\left[(p-3)\sqrt{p+1} \ \pm \ {8\,\epsilon\,\sqrt{2}}\,\sqrt{8-p}\right]}  \ ,
\eea
so that, in the parametrization of eqs.~\eqref{param_theta},
\beq{}{}{}{}{}
\sin\theta \ = \  \frac{3\,\epsilon\,(7-p)\sqrt{p+1}}{\left[(p-3)\sqrt{p+1} \ \pm \ {8\,\epsilon\,\sqrt{2}}\,\sqrt{8-p}\right]} \ .
\eeq
Different options are thus possible, in the two branches of these solutions, for the length of the internal interval and for the behavior of the string coupling. 

On the other hand, for $\phi_1=0$ the solutions in this class reduce to the supersymmetric ones obtained in Section~\ref{sec:killing_spinors} from the conditions determining the Killing spinors. In terms of the proper length, their limiting behavior is captured by eqs.~\eqref{universal_0limit}, which are not of the flux--free form~\eqref{param_theta}.

\item[3. ] For $E<0$ the range of $r$ is finite, together with the length of the interval, since the behavior at the two ends is the same as in eqs.~\eqref{universal}. Moreover, depending on the sign of $\beta_p$, the string coupling can vanish or diverge at both ends. It vanishes for RR forms for $p<3$, and also for the NS 7--form, while it blows up for $RR$ forms for $p>3$ and for the fundamental string. The effective Planck mass is finite in all cases. All these solutions break supersymmetry and have no counterparts in the absence of fluxes.
\end{itemize}

Summarizing, the $r$ interval can have a finite or infinite length, depending on the values of $C_1$ and $\phi_1$, the two constants in eqs.~\eqref{CphiZh}, and the string coupling diverges at least at one end, in most cases. However, due to the presence of fluxes, in some cases with broken supersymmetry the string coupling can remain bounded everywhere, even within intervals of finite length, and the effective Planck mass in $p+1$ dimensions can also be finite. 

\subsubsection{\sc Behavior of a Probe $D_p$-Brane} \label{sec:probe_brane}

One can describe a probe $p$-brane moving along the $r$ direction in the class of backgrounds of interest, considering the action
\beq{}{}
{\cal S} \ = \ - T_p V_p \int dt \ e^{\,-\,\phi}\,e^{A(p+1)}\sqrt{1 \ - \ e^{2(B-A)}\,\dot{r}^2 } \ + \ q_p V_p \ \int dt \ b_{p+1} \ .
\eeq
It can be deduced from the string--frame results of eq.~\eqref{t0hn0_gen_string}, so that
\beq{}{}{}
b'_{p+1} \ = \ \frac{H_{p+2}}{\left[f(r)\right]^2} \ .
\eeq
One is thus led, for $E>0$, to
\beq{}{}{}
b_{p+1} \ = \ - \ \frac{\sqrt{E}}{H_{p+2}}\ \coth\left(\sqrt{E} r\right) \ ,
\eeq
while the other cases with $E \leq 0$ can be obtained as a limit or by an analytic continuation,
and consequently
\beq{}{}
{\cal S} \ = \ - T_p V_p \int \frac{dt}{f(r)} \ \sqrt{1 \ - \ \dot{r}^2 f(r)\, e^{\frac{(3-p)p}{2(p+1)}\left(\phi_1 r+ \phi_2\right)}} \ - \ \frac{q_p V_p \sqrt{E}}{H_{p+2}} \ \int dt \ \coth\left(\sqrt{E} r\right) \ ,
\eeq
where $V_p$ denotes the volume in the spatial directions for the brane.
It is interesting to take a closer look at the behavior near $r=0$, where the background approaches the supersymmetric limit. For a non--relativistic brane
\beq{}{}
{\cal S} \ \simeq \ \int dt \left[ \ - \ \frac{T_p\,V_p}{f(r)} \ + \ \dot{r}^2\, \frac{T_p\,V_p}{2}\,e^{\frac{(3-p)p}{2(p+1)}\left(\phi_1 r+ \phi_2\right)} \right]  \ - \ \frac{q_p V_p \sqrt{E}}{H_{p+2}} \ \int dt \ \coth\left(\sqrt{E} r\right) \ ,
\eeq
and close to $r=0$ one can thus identify the potential
\beq{}{}{}
V(r) \ = \ \frac{V_p}{H_{p+2}\,r} \left(T_p+ q_p\right) \ .
\eeq
Its gravitational portion, proportional to the tension $T_p$, is repulsive, while its ``electric'' portion, proportional to the charge $q_p$, is also repulsive for $q_p>0$. Therefore, \emph{the singularity at $r=0$ behaves like a BPS extended object of negative tension and positive charge}, which in ten dimensions is naturally regarded as an orientifold, since the two contributions are equal in magnitude. While we have derived them working for $E>0$, these results apply for all values of $E$. Moreover, similar considerations apply to eleven--dimensional supergravity, where there is no dilaton and, as we have seen, $E>0$.

\subsection{\sc Solutions with $h$--Fluxes}

We can now turn to the second option that we have discussed in Section~\ref{sec:symmetries}, the presence of $h$--fluxes, with field strength
\beq{}{}{}{}
{\cal H}_{p+1} \ = \ h_{p+1} \, \epsilon_{p+1} \ , 
\eeq
with a constant $h_{p+1}$.

For non-selfdual cases, the starting point is now the system
 \bea
 A'' \!\!\!&=&  \frac{(D-p-2)}{2\,(D-2)} \ e^{2\,Y} h_{p+1}^2  \nonumber \  ,  \\
C''\!\!\!&=& - \  \frac{p}{2\,(D-2)}\ e^{2\,Y} h_{p+1}^2 \ ,  \nonumber  \\
   \phi'' \!\!\!&=& \frac{\beta_{p-1}\,(D-2)}{8}\ e^{2\,Y} h_{p+1}^2\ ,
  \eea
  which a special case of eqs.~\eqref{EqA_red}--\eqref{EqB_red}, where $p>0$ and
  \beq{}{}{}
  Y \ = \ (D-p-2)C\,-\,\beta_{p-1}\,\phi  \ ,
  \eeq
  after using the harmonic gauge condition. One can therefore work with the equation for $Y$,
 \beq{}{}{}
  Y'' \ = \ - \ {\widetilde{\Delta}}^2\, e^{2\,Y} \ ,
  \eeq 
  where
  \beq{}{}{}
  \widetilde{\Delta}^2  \ = \  \frac{{\widetilde{\delta}}^{\,2}}{{2}}\, h_{p+1}^2 \ ,
  \eeq
  and
  \beq{}{}{}
  \widetilde{\delta}^{\,2} \ = \ \frac{p(D-p-2)}{(D-2)} \ + \ \frac{\beta_{p-1}^2 \, (D-2)}{4} \ ,  \label{delta2}
  \eeq
which determines the behavior of $A$, $C$ and $\phi$, according to
 \bea{}{}
  A &=& - \ \frac{D-p-2}{\left(D-2\right) \widetilde{\delta}^{\,2}}\, Y \, +\, A_1\,r \,+\, A_2 \ , \nonumber \\ 
  \phi &=& - \  \frac{\beta_{p-1}\left(D-2\right)}{4\,\widetilde{\delta}^{\,2}} \,Y \, +\, \phi_1\,r \,+\, \phi_2 \ , \nonumber \\ 
  C &=& \frac{p}{\left(D-2\right){\widetilde{\delta}}^2}\,Y \ + \ \beta_{p-1}\ \frac{\phi_1\,r\,+\,\phi_2}{D-p-2} \ ,
  \eea
 where $A_1$, $A_2$, $\phi_1$ and $\phi_2$ are constants.
The equation for $Y$ can turned into the non--relativistic energy conservation condition
  \beq{}{}{}
  \left(Y'\right)^2 \ = \ - \ \widetilde{\Delta}^2\, e^{2\,Y} \ + \ \widetilde{E} \ ,
  \eeq
where now $\widetilde{E}$ must be \emph{positive}, while the Hamiltonian constraint reads
 \begin{align}
&(p+1)A'[p\,A' \,+\, (D-p-2)C']\ +\ (D-p-2)C'[(D-p-3)C'+(p+1)A'] \nonumber \\
 &- \ \frac{4\,(\phi')^2}{D-2}  \ -\ \frac{1}{2}\, e^{\,2\,Y} \ h_{p+1}^2  \,= \, 0 \ , \label{EqB_red_h}
 \end{align}
and allows to relate $\widetilde{E}$ to the integration constants, according to
\beq{}{}{}
\widetilde{E} \ = \ \widetilde{\delta}^{\,2} \left[ p(p+1) \left(A_1 \,+\,\frac{\beta_{p-1}}{p}\, \phi_1\right)^2 \ - \ \phi_1^2\left(\frac{\beta_{p-1}^2\left(D-2\right)}{p\left(D-p-2\right)} \ + \ \frac{4}{D-2}\right) \right] \ .
\eeq
Equivalently
\beq{}{}{}
\widetilde{E} \ = \ \widetilde{\delta}^{\,2} \left[ p(p+1) \left(A_1 \,+\,\frac{\beta_{p-1}}{p}\, \phi_1\right)^2 \ - \ \frac{4\,\phi_1^2\,\widetilde{\delta}^{\,2}}{p\left(D-p-2\right)} \right] \ ,
\eeq
where we used eq.~\eqref{delta2}.
This result is consistent with eq.~\eqref{energy_h} and with the symmetry transformation in eq.~\eqref{sym_AC}. 

In ten dimensions, after using the Ramond--Ramond expression for $\beta_{p-1}$, letting $\widetilde{E} = \frac{1}{\widetilde{\rho}^2}$, the preceding expression can be cast in the form
\beq{}{}
1 \ = \ 2 \, \widetilde{\rho}^2 \left[ p(p+1) \left(A_1 \,+\,\frac{p-4}{4\,p}\, \phi_1\right)^2 \ - \ \frac{8\,\phi_1^2}{p\left(8-p\right)} \right] \ ,
\eeq
and one can parametrize the solutions according to
\bea{}{}
\phi_1 &=& \frac{1}{4 \widetilde{\rho}}\,\sqrt{p(8-p)}\, \sinh \widetilde{\zeta} \ , \nonumber \\
A_1 &=& - \ \frac{p-4}{16\,\widetilde{\rho}}\, \sqrt{\frac{8-p}{p}}\,\sinh\widetilde{\zeta} \ \pm \ \frac{1}{\widetilde{\rho} \sqrt{2 p(p+1)}}\, \cosh \widetilde{\zeta} \ .
\eea

According to Appendix~\ref{app:deq},
\beq{}{}{}
e^{\,-\,Y} \ \equiv \ {\widetilde{f}}(r) \ = \ \left|h_{p+1}\right| \, \cosh \left(\frac{r}{\widetilde{\rho}}\right) \ ,
\eeq
so that $- \infty < r < +\infty$, and consequently
\bea
ds^2 &=& \left[{\widetilde{f}}(r)\right]^{\,\frac{8-p}{8}} \left[ e^{\,2\left(A_1\,r\,+\,A_2\right)}\,dx^2 \ + \ e^{\,\frac{p-4}{2} (\phi_1\,r+\phi_2)+2(p+1)(A_1\,r+A_2)}dr^2 \right] \nonumber \\ &+& \left[{\widetilde{f}}(r)\right]^{\,-\,\frac{p}{8}} \, e^{\,\frac{\left(p-4\right)}{2(8-p)}(\phi_1\,r+\phi_2)}\,d\vec{y}^{\,2} \ , \nonumber \\
e^\phi &=& \left[{\widetilde{f}}(r)\right]^{\,\frac{p-4}{4}}\, e^{\, \phi_1\,r+\phi_2}\ , \nonumber \\
{\cal H}_{p+1} &=& h_{p+1}\, \epsilon_{p+1} \ . \label{gen_sol_hnoT_h}
\eea
In this setting the string coupling can be bounded everywhere provided
\beq{}{}{}
\pm \sqrt{p(8-p)}\,\sinh \widetilde{\zeta} \ + \ p \ - \ 4 \,<\, 0 \ ,
\eeq
or
\beq{}{}{}
p \ < \ 4 \ , \qquad \left|\sinh\widetilde{\zeta} \right| \ \leq \ \left|\frac{p-4}{\sqrt{p(8-p)} }\right| \ ,
\eeq
but the length of the internal interval and the effective Planck mass are always infinite.
There are again several options for the length of the internal interval and for the effective Planck mass, and the $p=4$ case must treated separately, since the five form in type IIB is self-dual.

\section{\sc Cosmologies with Fluxes} \label{sec:cosmoflux}

There are also interesting cosmological counterparts of the preceding solutions, which can be obtained by the analytic continuation of $r$ to $\tau$ in the original system of eqs.~\eqref{CphiZ_space}. 

\subsection{\sc Cosmological Solutions with $H$--Fluxes}

The analytic continuation has a notable consequence: it reverts the Newtonian potential, so that the solutions are related to the case $\epsilon=-1$ in Appendix~\ref{app:deq}, and rest on 
\beq
g(\tau) \ = \ \left|H_{p+2}\right|\,\rho\,\cosh\left( \frac{\tau}{\rho} \right) \ .
\eeq
As a result, these cosmologies are described by
\bea
ds^2 &=& \left[g(\tau)\right]^{\,\frac{p-7}{8}} \, e^{\,-\,\frac{2\,\beta_p(\phi_1\,\tau+\phi_2)}{p+1}}dx^2 \ - \ \left[g(\tau)\right]^{\frac{p+1}{8}}  \, e^{\,-\,{2\,\beta_p (\phi_1\,\tau+\phi_2)}+2(8-p)(C_1\,\tau+C_2)}d\tau^2 \nonumber \\ &+& \left[g(\tau)\right]^{\frac{p+1}{8}} \, e^{\,2(C_1\,\tau+C_2)}d\vec{y}^{\,2} \ , \nonumber \\
e^\phi &=& \left[g(\tau)\right]^{\,-\,\beta_p}\, e^{\,2\, (\phi_1\,\tau+\phi_2)}\ , \nonumber \\
{\cal H}_{p+2} &=& H_{p+2}\, \frac{\epsilon_{p+1}\, dr}{\left[g(\tau)\right]^2} \ .
\eea
Here $ - \infty < \tau <\infty$, and there are two branches for $C_1$ and $\phi_1$, which can be parametrized as   
    \bea
    C_1 &=& \frac{\beta_p}{4\,\rho} \, \sqrt{\frac{p+1}{7-p}}\, \sinh \zeta \ \pm \ \frac{\cosh \zeta}{\rho\,\sqrt{2(8-p)(7-p)}} \ , \nonumber \\
    \phi_1 &=& \frac{\sqrt{(p+1)(7-p)}}{4\,\rho}\, \sinh \zeta \ , \label{C1f1Eposc}
    \eea
    for $p<7$, and as
    \beq
    \phi_1 \ = \ \pm \ \frac{2}{\sqrt{3}\,\rho} \ e^{\,\pm\,\zeta} \ , \qquad
    C_1 \ = \ \frac{\sqrt{3}}{4\,\rho}\, \sinh \zeta  \label{C1f1Epos7c}
    \eeq
    for $p=7$.
    
    Different types of asymptotic behavior are possible as $\tau \to \epsilon \,\infty$, where $\epsilon=\pm 1$, as can be seen from the limiting forms,
\bea
ds^2 &\sim& - \ e^{\,2\,b\,\tau}\, d\tau^2 \ + \ e^{\,2\,a\,\tau}\, dx^2  \ + \ e^{\,2\,c\,\tau}\, dy^2 \ , \nonumber \\
e^\phi &\sim& e^{\,f\,\tau} \ ,
\eea
with
\bea
a &=& - \ \frac{1}{\rho} \left[\frac{7-p}{16}\,\epsilon \ + \ \frac{\beta_p}{4} \, \sqrt{\frac{7-p}{p+1}}\, \sinh \zeta \right]  \ , \nonumber \\
b &=&  \frac{1}{\rho} \left[\frac{p+1}{16}\,\epsilon \ + \ \frac{\beta_p}{4}\, \sqrt{\frac{p+1}{7-p}}\, \sinh\zeta \ \pm \ \sqrt{\frac{8-p}{2(7-p)}}\, \cosh \zeta\right] \ , \nonumber \\
c &=& \frac{1}{\rho} \left[\frac{p+1}{16}\,\epsilon \ + \ \frac{\beta_p}{4}\, \sqrt{\frac{p+1}{7-p}}\, \sinh\zeta \ \pm \ \frac{1}{\sqrt{2(7-p)(8-p)}}\, \cosh \zeta\right]\ , \nonumber \\
f &=&  \frac{1}{\rho} \left[ \ - \ \beta_p\,\epsilon \ + \ \frac{1}{4}\ \sqrt{(p+1)(7-p)}\, \sinh\,\zeta\right]\ ,  \label{large_r_asymptc}
\eea
while for $p=7$, with the parametrization~\eqref{C1f1Epos7c}, the coefficients become
\bea
a &=& \mp \ \frac{e^{\pm \zeta}}{4\sqrt{3}\,\rho} \ , \nonumber \\
b &=& \frac{1}{\rho} \left[ \frac{\epsilon}{2} \ \mp \ \frac{2\ e^{\pm \zeta}}{\sqrt{3}} \ + \ \frac{\sqrt{3}}{4} \ \sinh \zeta \right] \ , \nonumber \\
c &=& \frac{1}{\rho} \left[ \frac{\epsilon}{2} \ + \ \frac{\sqrt{3}}{4} \ \sinh \zeta \right] \ , \nonumber \\
f &=& \frac{1}{\rho} \left[  \pm \ \frac{2\ e^{\pm \zeta}}{\sqrt{3}} \ - \ 1 \right]  \label{large_r_asympt7c} \ .
\eea
The solutions manifest a power--like behavior in terms of the cosmic time for large values of $\tau$, close to the initial and final singularities, with different scale factors for the physical and internal portions of the space: 
\beq{}{}
ds^2 \ = \ - \ dt^2 \ + \ a_1^2(t)\, dx^2 \ + \ a_2^2(t)\, dy^2 \ .
\eeq
Here
\beq{}{}
a_1 \ \sim \ \left|b \,t\right|^\frac{a}{b}  \ , \qquad a_2 \ \sim \ \left|b\,t\right|^\frac{c}{b} \ ,
\eeq
so that different types of scenarios are possible, with both portions of the space expanding or one expanding and one contracting, depending on the relative signs of these quantities. This is also true for the string coupling, which can attain strong or weak coupling at the singularities. In the RR case the string coupling can be bounded during the whole cosmological evolution if $p \geq 3$ and $|\zeta|$ is small enough, and the same is true in the NS-NS case for $p=1$.

\subsection{\sc Cosmological Solutions with $h$--Fluxes}

The solutions for this case can be deduced from eqs.~\eqref{gen_sol_hnoTg}, making use the transformations of eqs.\eqref{sym_AC}. One thus obtains
\bea
ds^2 &=& \left[f(\tau)\right]^{\,2\,\sigma_B} \left[ e^{\,2(A_1\,r+A_2)}\,d\vec{x}^{\,2}  \ - \ e^{\,{2\,\beta_{p-1} (\phi_1\,r+\phi_2)}+2(p+1)(A_1\,r+A_2)}d\tau^2\right] \nonumber \\ &+& \left[f(\tau)\right]^{\,-\,2\,\sigma_A} \, e^{\,\frac{2\,\beta_{p-1}\left(\phi_1\,\tau\,+\,\phi_2\right)}{D-p-2}}d\vec{y}^{\,2} \ , \nonumber \\
e^\phi &=& \left[f(\tau)\right]^{\,-\,\sigma_\phi}\, e^{\, \phi_1\,\tau+\phi_2}\ , \nonumber \\
{\cal H}_{p+1} &=& h_{p+1}\, {\epsilon_{p+1}}\ , \label{gen_sol_hnoTgh}
\eea
where
\beq
\sigma_A \,=\, \frac{p}{\widetilde{\delta}^2\,(D-2)} \ , \qquad
\sigma_B \,=\,  \frac{(D-p-2)}{\widetilde{\delta}^2\,(D-2)}\ , \qquad \sigma_\phi \,=\,-\, \frac{\beta_{p-1}\,(D-2)}{4\,\widetilde{\delta}^2} \ ,
\eeq
and
\beq
\widetilde{\delta}^2 \ = \ \frac{1}{(D-2)}\left[\left(\frac{\beta_{p-1}\,(D-2)}{2}\right)^2 \,+\, p(D-p-2) \right] \ .
\eeq
Here
\beq{}{}
f(\tau) \ = \ \frac{\widetilde{\Delta}}{\sqrt{E}}\, \sinh\left(\sqrt{E} \tau\right) \ , \label{fEtau2}
\eeq
\beq
\widetilde{\Delta}^2 \ = \ \frac{h_{p+1}^2\, \widetilde{\delta}^2}{2}   \ .
\eeq
and
\beq
E \ = \ \widetilde{\delta}^2\,\Bigg[ \left(\frac{(D-p-3)\,\beta_{p-1}^2}{D-p-2} \ - \ \frac{4}{D-2} \right)\phi_1^2 \ + \ (p+1)A_1 \left[ p \,A_1 \ + \ 2\,\beta_{p-1}\,\phi_1\right]\Bigg] \ . \label{energy_hh}
\eeq
There are now two cases of interest. The first is $D=11$, where $\beta_{p-1}=0$ and $p=3,6$. The second is $D=10$, where in the RR case $p$ is even in Type IIA and odd in Type IIB and $\beta_{p-1}=\frac{p-4}{4}$. The $p=4$ case is special, since the form is self--dual, and must be treated separately. On the other hand, in the NS-NS case $p=2$ with $\beta_{1}=\frac{1}{2}$ and $p=6$ with $\beta_{5}=-\,\frac{1}{2}$, as can be seen in Table~\ref{table:tab_2}.

\subsubsection{\sc The $D=11$ Cases}
In these cases $\widetilde{\delta}^2=2$, and one can see from eq.~\eqref{energy_hh} that $E \geq 0$. Therefore, for $p=3$ one obtains the five--dimensional cosmologies
\bea
ds^2 &=& \left[f(\tau)\right]^{\,\frac{2}{3}} \left[ e^{\,2(A_1\,r+A_2)}\,d\vec{x}^{\,2}  \ - \ e^{\,8(A_1\,r+A_2)}\,d\tau^2 \right] \nonumber \,+\,  \left[f(\tau)\right]^{\,-\,\frac{1}{3}} \, d\vec{y}^2 \ , \nonumber \\
{\cal H}_{4} &=& h_{4}\, {\epsilon_{4}}\ , \qquad f(\tau) \ = \ \frac{h_4}{2\sqrt{6}\left|A_1\right|} \ \sinh\left(2\sqrt{6}\left|A_1\right| \tau\right) \ , \label{gen_sol_hnoTgh11h1}
\eea
while for $p=6$ one obtains four--dimensional cosmologies if the $x$ are compact coordinates  and the $y$ are non--compact coordinates 
\bea
ds^2 &=& \left[f(\tau)\right]^{\,\frac{1}{3}} \left[ e^{\,2(A_1\,\tau+A_2)}\,d\vec{x}^{\,2}  \ - \ e^{\,14(A_1\,\tau+A_2)}\,d\tau^2\right] \,+\, \left[f(\tau)\right]^{\,-\,\frac{2}{3}} \,d\vec{y}^{\,2} \ , \nonumber \\
{\cal H}_{4} &=& h_{7}\, \frac{{\epsilon}_{3}\,d\tau}{\left[f\left(\tau\right)\right]^2}\ , \qquad f(\tau) \ = \ \frac{h_7}{2\sqrt{21}\left|A_1\right|} \ \sinh\left(2\sqrt{21}\left|A_1\right| \tau\right) \ ,
\eea
 but in this case one must let $\tau$ run backwards from $+ \infty$.
The $E=0$ solutions can be recovered in the limit $A_1\to 0$.
\subsubsection{\sc The $D=10$ Cases}

In all these cases (RR for $p \neq 4$ and NS-NS) there are three classes of solutions, as in Section~\ref{h_flux_sols}, depending on the value of $E$.
\begin{itemize}
\item[1. ] If $ E > 0$ in eq.~\eqref{E_value}, letting $E = \frac{1}{\rho^2}$, one can see from eq.~\eqref{fEtau2} that
\beq
f(\tau) \ = \ \widetilde{\Delta}\,\rho\,\sinh\left(\frac{\tau}{\rho}\right) \ , \label{fposEh}
\eeq
where $0 < \tau < \infty$, up to a translation and a choice of orientation on the $r$ axis. Eq.~\eqref{energy_hh} then shows that, for $p>0$, the independent choices for $A_1$ and $\phi_1$ can be expressed in terms of a real parameter $\zeta$ according to
    \bea
    A_1 &=& -\,\frac{\beta_{p-1}}{4\,\rho} \, \sqrt{\frac{8-p}{p}}\, \sinh \zeta \ \pm \ \frac{\cosh \zeta}{\rho\,\sqrt{2p(p+1)}} \ , \nonumber \\
    \phi_1 &=& \frac{\sqrt{p(8-p)}}{4\,\rho}\, \sinh \zeta \ , \label{C1f1Eposh}
    \eea
    while for $p=0$
    \beq
    \phi_1 \ = \ \pm \ \frac{2}{\sqrt{3}\,\rho} \ e^{\,\pm\,\zeta} \ , \qquad
    A_1 \ = \ \frac{\sqrt{3}}{4\,\rho}\, \sinh \zeta \ . \label{C1f1Epos0h}
    \eeq

\item[2. ] If $E=0$, one can see from one can see from eq.~\eqref{fEtau2}
\beq
f(\tau) \ = \ \widetilde{\Delta}\,\tau \ ,
\eeq
where $0 < \tau < \infty$, and then for $p>0$
\beq
A_1 \ = \ \frac{\phi_1}{p} \left[ -\,\beta_{p-1} \ \pm \ \frac{2\,\sqrt{2}\,\sign(\phi_1)}{\sqrt{(p+1)(8-p)}}\right] \ ,
\label{one_param_E0h}
\eeq
while for $p=0$ there are again two branches of solutions, with
    \beq
    \phi_1 \ = \ 0 \label{p7E0firsth}
    \eeq
    and $A_1$ arbitrary, which are Kasner--like, or with
    \beq
    \phi_1 \ = \ \frac{16}{3}\, A_1 \ . \label{p7E0secondh}
    \eeq

\item[3. ] If $E<0$, letting $E = \,-\,\frac{1}{\rho^2}$, one can see from eq.~\eqref{fEtau2} that
\beq
f(\tau) \ = \ \widetilde{\Delta}\,\rho\,\sin\left(\frac{\tau}{\rho}\right) \ ,
\eeq
where $0 < \tau < \pi\,\rho$, up to a translation and a choice of orientation on the $\tau$ axis. Eq.~\eqref{energy_hh} then shows that for $p>0$ the independent choices for $A_1$ and $\phi_1$ can be expressed in terms of a real parameter $\zeta$, according to
\bea
\phi_1 &=& \pm \ \frac{\sqrt{p(8-p)}}{4\,\rho}\, \cosh \zeta \ , \nonumber \\
 A_1 &=& \mp\ \frac{\beta_{p-1}}{4\,\rho} \, \sqrt{\frac{8-p}{p}}\, \cosh \zeta \ + \ \frac{\sinh \zeta}{\rho\,\sqrt{2 p (p+1)}} \ . \label{paramEnegh}
\eea
On the other hand, for $p=7$ there are again two branches of solutions, with
    \beq
    \phi_1 \ = \ \pm \ \frac{2}{\sqrt{3}\,\rho} \ e^{\,\pm\,\zeta} \ , \qquad
    A_1 \ = \ \frac{\sqrt{3}}{4\,\rho}\, \sinh \zeta \ . \label{paramEneg7h}
    \eeq
\end{itemize}

Within this class of solution it is relatively simple to attain a bounded string coupling throughout a four--dimensional cosmological evolution. This is the case, for example, for the upper branch of four--dimensional solutions in eqs.~\eqref{C1f1Eposh} with an NS-NS flux ($p=2, \beta_{1}=\frac{3}{4}$).

%%%%%%%%%%%%%%%%%%%%%%%%%%%%%%%%%%
\section{\sc  Conclusions}\label{sec:conclusions}
%%%%%%%%%%%%%%%%%%%%%%%%%%%%%%%%%%%%%
\vskip 12pt

In this paper, starting from the special class of metric profiles of eq.~\eqref{metric_sym_intro}, we have set up a formalism that can encompass a wide class of compactifications (or cosmologies) depending on a single coordinate. These involve, in general, a maximally symmetric spacetime and a maximally symmetric internal space, and are also suitable to describe brane profiles.  Here we have confined our attention to compactifications from $D=10$ and $D=11$ to Minkowski spaces, resorting to a simple flat internal manifold, a torus, but one can conceive generalizations of our results that involve Ricci--flat internal spaces. A warping factor is always present, and plays an important role in these types of vacuum solutions. Settings of this type induce a spontaneous breaking of supersymmetry, in ways that are controlled by tunable parameters. 

After discussing the supersymmetric options that, aside from flat space, require the presence of internal fluxes, we have considered vacuum solutions that break supersymmetry in the absence of fluxes. These are generalized Kasner solutions that are characterized by a singularity at a finite distance, and the string coupling diverges either there or at infinity. We then examined the general effects of a form flux compatible with the symmetries of the metrics in eq.~\eqref{metric_sym_intro}. This addition resulted in the emergence of three families of backgrounds, which are distinguished by the energy $E$ of a Newtonian particle subject to an exponential potential. The three families of solutions have in common a singular boundary at the origin, where the form field strength diverges and where they approach a supersymmetric limit, which belongs to the zero--energy family. The behavior of a probe brane in the vicinity of this boundary revealed the presence  at the origin of a BPS--like extended object, with negative tension and positive charge. In the resulting allowed region, the two families of non--supersymmetric solutions with $E \geq 0$ interpolate between a supersymmetric vacuum at the origin and one of the Kasner--like vacua with no flux at the other boundary, which can lie at a finite or infinite distance from it. The solutions of the third family, with $E<0$, have a second boundary at a finite distance from the origin where they approach again a supersymmetric behavior.
Curvature singularities remain present in all these vacua~\cite{cond_dud}, but the string coupling is everywhere bounded in some cases. We have characterized precisely their moduli spaces, which include special subsets of parameters where supersymmetry is recovered globally.  We have also studied corresponding classes of cosmological models, which can be obtained from the vacuum solutions via analytic continuations. In one class of these models, it is simple and natural to obtain finite values of the string coupling throughout the cosmological evolution.

The solutions that we have discussed result in lower--dimensional Minkowski spaces, and in this respect they are somewhat reminiscent of the well-known and widely explored Calabi-Yau vacua~\cite{strings}, which led to systematic investigations of the partial breaking of supersymmetry in String Theory. Ending up in Minkowski space is clearly interesting in connection with Particle Physics. However, it also hides internal scales, opening a potentially useful window for perturbatively stable vacua. This feature should be contrasted with the behavior of the non--supersymmetric $AdS \times S$ solutions of~\cite{nonsusyads}, where the scales of the internal spheres and of the $AdS$ spacetimes are related, and consequently unstable Kaluza--Klein modes show up inevitably~\cite{bms} (see~\cite{review_17} and~\cite{review_21} for reviews). Among the vacua that we have analyzed, we have found a particularly interesting class of IIB compactifications to four dimensions supported by a flux of the self--dual five form, where the string coupling is everywhere bounded. In~\cite{int4d_vacuum} we shall explore further these fluxed compactifications to four dimensions, addressing in detail the key issue of perturbative stability, which has proved daunting for non--supersymmetric $AdS$ vacua. 

A common feature of non--supersymmetric vacua is the emergence of severe back-reactions. This will be the case for all the examples discussed in this paper, once quantum effects are taken into account. Back-reactions have long been recognized as an important issue in the widely explored context of Scherk-Schwarz compactifications~\cite{scherkschwarz}, which have simple realizations in String Theory~\cite{scherkschwarz_closed}, even when open string are present~\cite{scherkschwarz_open}. However, similar back-reactions present themselves in a more basic context already in ten dimensions, in the three ten--dimensional non--supersymmetric non--tachyonic models of~\cite{so16so16}, \cite{susy95}, and of~\cite{bsb}, where supersymmetry is non-linearly realized~\cite{dmnonlinear}, and lead to the emergence of a peculiar contribution to the low--energy effective action, a  ``tadpole potential''
\beq{}
\Delta\,{\cal S} \ = \ - \ T \int d^{10} x \, \sqrt{-g}\ e^{\gamma \phi} \ , \label{tadpole_pot}
\eeq
where $\gamma=\gamma_c=\frac{3}{2}$ for the orientifold models and $\gamma=\frac{5}{2}$ for the heterotic $SO(16)\times SO(16)$ model. The same types of potentials are the leading contributions to generic string models when quantum corrections due to broken supersymmetry are taken into account.
Their effects on vacuum solutions of the type considered here will lead to generalizations of the Dudas--Mourad vacuum~\cite{dm_vacuum}, of their counterparts for generic values of $\gamma$ in~\cite{russo}, and of the climbing--scalar cosmologies~\cite{dks}, with a number of novel features. These systems are the main topic of the companion paper~\cite{ms_vacuum_2}.

\vskip 12pt
%%%%%%%%%%%%%%%%%%%%%%%%%%%%%%
\section*{\sc Acknowledgments}
%%%%%%%%%%%%%%%%%%%%%%%%%%%%%%

\vskip 12pt

We are grateful to E.~Dudas for stimulating discussions, and also to G.~Bogna and Y.~Tatsuta, who read an earlier version of the manuscript and made useful comments. The work of AS was supported in part by Scuola Normale, by INFN (IS GSS-Pi) and by the MIUR-PRIN contract 2017CC72MK\_003. JM is grateful to Scuola Normale Superiore for the kind hospitality extended to him while this work was in progress. AS is grateful to U. Paris VII and DESY--Hamburg for the kind hospitality, and to the Alexander von Humboldt foundation for the kind and generous support, while this work was in progress.

%%%%%%%%%%%%%%%%%%%%%%%%%%%%%%%%%%%%%%%%%%%%%%%%%
\vskip 24pt

\vskip 24pt
\newpage
\appendix
\begin{appendices}

\section{\sc Some Technical Details} \label{app:technical}

For the backgrounds of eq.~\eqref{metric_sym_intro}, the Riemann tensor is computed most conveniently starting from the one--form spin connection, with non--vanishing components
\beq
{\omega^\mu}_r \ = \ - \ {\omega_r}^\mu \ =\ e^{A-B}\,A'\,dx^\mu \ , \qquad {\omega^i}_r \ = \ - \ {\omega_r}^i \ = \ e^{C-B}\,C'\,dy^i \ . 
\eeq
The non--vanishing contributions to the curvature two-form
\beq
{\Omega_A}^B  \ \equiv \ \frac{1}{2}\, {{R_{MN}}_A}^{\ \ B} \ dx^M\wedge dx^N \ = \ d\, {\omega_A}^B \ + \ {\omega_A}^C\wedge {\omega_C}^B \ ,
\eeq
when converted to curved indices, determine the background Riemann tensor, where here we add also the contributions from the curvatures of the maximally symmetric metrics  $\gamma_{\mu\nu}$ and $\gamma_{mn}$:
\bea
{\Omega}^{r \mu} &=& - \ \left( e^{A-B} \, A'\right)'\, dr \wedge  dx^\mu \ , \qquad {\Omega}^{r m} \ = \ - \ \left( e^{C-B} \, C'\right)'\, dr\wedge dy^m \ , \nonumber \\
{\Omega}^{\mu\nu} &=& - \ \left[e^{2(A-B)} \left(A'\right)^2\,-\,\frac{k}{\alpha'} \right] dx^\mu \wedge dx^\nu \ , \nonumber \\
{\Omega}^{m\nu} &=& - \ e^{A+C-2B}\,A'\,C'\,dy^m \wedge dx^\nu \ , \nonumber \\ 
{\Omega}^{mn} &=&  - \ \left[e^{2(C-B)} \left(C'\right)^2\,-\,\frac{k'}{\alpha'} \right] dy^m \wedge dy^n \ .
\eea
Consequently, the Riemann tensor with all curved indices has the non--vanishing components
\bea
{R_{r\mu r}}^{\nu} &=& - \ {\delta_\mu}^\nu \ e^{B-A}\left( e^{A-B} \, A'\right)' \ , \qquad {R_{r m r}}^{n} \ = \ - \ {\delta_m}^n \ e^{B-C}\left( e^{C-B} \, C'\right)' \ , \nonumber \\
{R_{\alpha\beta\mu}}^{\nu} &=& - \ \gamma_{\mu[\alpha}\, {\delta_{\beta]}}^\nu \left[ e^{2(A-B)} \left(A'\right)^2\,-\,\frac{k}{\alpha'}\right] \ , \qquad
{R_{\nu n \mu}}^{m} \ = \ - \ e^{2(A-B)}\,\gamma_{\mu\nu}\,{\delta_n}^m\,A'\,C' \ , \nonumber \\ {R_{mnp}}^{q} &=& - \ \gamma_{p[m}\, {\delta_{n]}}^q \left[ e^{2(C-B)} \left(C'\right)^2\,-\, \frac{k'}{\alpha'}\right]\ , \qquad {R_{\nu n m}}^{\mu} \ = \ e^{2(C-B)}\,{\delta_{\nu}}^\mu\,\gamma_{mn}\,A'\,C' \ ,
\eea
together with others obtained from these by symmetry, and thus
\bea
R_{\mu\nu}^{(0)} &=& - \ \gamma_{\mu\nu}e^{2A}\left[p(A')^2 \,e^{-2B} \ +\  (D-p-2)A'\,C'e^{-2B}\ +\ \left(A'e^{A-B}\right)'e^{-A-B} \right. \nonumber \\&-& \left. \frac{p \,k}{\alpha'} \,e^{-2A}\right]\ , \nonumber \\
R_{rr}^{(0)} &=& - \left[ (p+1)\left(A'e^{A-B}\right)'e^{B-A}\ + \ (D-p-2)\left(C'e^{C-B}\right)'e^{B-C}\right] \ , \nonumber \\
R_{mn}^{(0)} &=&  -\ \gamma_{mn}e^{2C}\left[(D-p-3)(C')^2 \,e^{-2B}\ + \ (p+1)A'\,C'e^{-2B} \ +\ \left(C'e^{C-B}\right)'e^{-B-C} \right. \nonumber \\&-& \left. \frac{(D-p-3) \,k'}{\alpha'} \,e^{-2C}\right]\ .
\eea
In the harmonic gauge~\eqref{harm_gauge}
\bea{}{}
R_{\mu\nu}^{(0)} &=& - \ \gamma_{\mu\nu}\left[ e^{2(A-B)} \, A'' \ - \ \frac{p\,k}{\alpha'} \right] \ , \nonumber \\
R_{rr}^{(0)} &=& - \Big[\left(p+1\right)A'' \,+\, \left(D-p-2\right)C''\,-\,p(p+1)\left(A'\right)^2 \nonumber \\
&-& (D-p-2)(D-p-3)\left(C'\right)^2 \,-\, 2(p+1)(D-p-2) A'C'\Big]\,, \nonumber \\
R_{mn}^{(0)} &=& - \ \gamma_{mn}\left[ e^{2(C-B)} \, C'' \ - \ \frac{(D-p-3)\,k'}{\alpha'} \right] \ .
\eea
Consequently the scalar curvature is
\bea{}{}{}
R &=& -2 e^{-2B}\left[(p+1) A'' \,+\,(D-p-2) C''\right]\nonumber \\ &+&\frac{p(p+1)k}{\alpha'}\,e^{-2A}\,+\,\frac{(D-p-1)(D-p-2)k'}{\alpha'}\,e^{-2C} \\
&+& e^{-2B}\left[p(p+1)(A')^2+2(p+1)(D-p-2) A'C'+(D-p-2)(D-p-3) (C')^2 \right] \, . \nonumber
\eea
and therefore
\bea{}{}{}
G_{rr} &=& - \ \frac{1}{2}\,\frac{p(p+1)k}{\alpha'}\,e^{2(B-A)}\,- \ \frac{1}{2}\,\frac{(D-p-1)(D-p-2)k'}{\alpha'}\,e^{2(B-C)}  \\
&+& \frac{1}{2}\left[p(p+1)(A')^2+2(p+1)(D-p-2) A'C'+(D-p-2)(D-p-3) (C')^2 \right] \,. \nonumber
\eea
In order to build the contributions of tensor fields strengths to the Einstein equations one needs to take into account some properties of the field strengths of eqs.~\eqref{profile_H} and \eqref{profile_h},
\bea
\left({{\cal H}_{p+2}}^2\right)_{\mu\nu} &=& -\ (p+1)! \left(b'{}_{p+1}(r)\right)^2 e^{-2 p A - 2 B} \ \gamma_{\mu\nu} \ , \nonumber \\
\left({{\cal H}_{p+2}}^2\right)_{rr} &=& -\ (p+1)! \left(b'{}_{p+1}(r)\right)^2 e^{-2 (p+1) A } \ ,
\eea
and
\beq
\left({{\cal H}_{p+1}}^2\right)_{\mu\nu} \ = \ -\ p! \ h_{p+1}^2 e^{-2 p A} \gamma_{\mu\nu} \ . \label{h2mn}
\eeq
There are also two special cases related to self--dual forms. The first concerns the
self--dual profile of eq.~\eqref{H5H}, for which
\bea
\left({{\cal H}_{5}}^2\right)_{\mu\nu} &=& -\ 3 \, H_5^2 \ e^{2 A - 10 C} \ \gamma_{\mu\nu} \ , \quad
\left({{\cal H}_{5}}^2\right)_{rr} \,=\, -\ 3  \,H_5^2 \ e^{2 B - 10 C } \ , \nonumber \\
 \left({{\cal H}_{5}}^2\right)_{mn} &=& 3 \, H_5^2 \ e^{ - 8 C} \ \gamma_{mn} \ ,
\eea
consistently with
\beq
{\cal H}_{5}^2 \ = \ 0 \ ,
\eeq
which is implied by the self--duality. The second concerns the corresponding $h$-profile of eq.~\eqref{H5h}, for which 
\bea
\left({{\cal H}_{5}}^2\right)_{\mu\nu} &=& -\ 3 \, h_5^2 \ e^{- 8 A} \ \gamma_{\mu\nu} \ , \quad
\left({{\cal H}_{5}}^2\right)_{rr} \,=\, 3 \, h_5^2 \ e^{2 B - 10 A } \ , \nonumber \\
 \left({{\cal H}_{5}}^2\right)_{mn} &=& 3 \,h_5^2 \ e^{2 C - 10 A} \ \gamma_{mn} \ . 
\eea
\section{\sc A Useful Newtonian Model} \label{app:deq}

In this Appendix we discuss the differential equation
 \beq
 Z'' \ = \ \epsilon\,\Delta^2 \ e^{\,2\,Z} \ , \label{eqZ}
 \eeq
 where $\epsilon=\pm 1$ and $\Delta$ is real and positive, which plays a role in various parts of the paper. This equation has the first integral
 \beq
 (Z')^2 \ = \ E \ + \ \epsilon\, \Delta^2\, e^{2\, Z} \ , \label{conserv}
 \eeq
 which has the form of an energy conservation condition for a Newtonian particle,
 and in order to proceed further one must distinguish a few cases.
 \begin{enumerate}

 \item If $\epsilon = 1$ and $E=\frac{1}{\rho^2}>0$, letting
 \beq
 x \ = \ \frac{r}{\rho}
 \eeq
 eq.~\eqref{conserv} becomes
\beq
 \left(\frac{d Z}{d x}\right)^2  \ = \ 1 \ + \ \left(\Delta\,\rho\right)^2 \, e^{2\, Z} \ .
\eeq
The redefinition
\beq
Z = \ \tilde{Z} - \log(\Delta\, \rho)
\eeq
leads to the reduced equation
\beq
 \left(\frac{d \tilde{Z}}{d x}\right)^2  \ = \ 1 \ + \ e^{2\, \tilde{Z}} \ ,
\eeq
and therefore finally to the solutions
\beq
Z \ = \ -  \log\left[\Delta\,\rho \, \sinh \left(\frac{r-r_0}{\rho}\right) \right]\ ,
\eeq
in the region $r > r_0$, or
\beq
Z \ = \ -  \log\left[\Delta\,\rho \,\sinh\left(\frac{r_ 0 - r}{\rho}\right) \right]\ ,
\eeq
in the region $r < r_0$.
\item If $\epsilon = 1$ and $E=- \frac{1}{\rho^2} < 0$, letting
\beq
 x \ = \ \frac{r}{\rho}
 \eeq
 eq.~\eqref{conserv} becomes
\beq
 \left(\frac{d Z}{d x}\right)^2  \ = \ - \,1 \ + \ \left(\Delta\,\rho\right)^2 \, e^{2\, Z} \ .
\eeq
The redefinition
\beq
Z = \ \tilde{Z} - \log(\Delta\, \rho)
\eeq
leads to the reduced equation
\beq
 \left(\frac{d \tilde{Z}}{d x}\right)^2  \ = \ - \,1 \ + \ e^{2\, \tilde{Z}} \ ,
\eeq
and therefore finally to the solution
\beq
Z \ = \ -  \log\left[\Delta\,\rho \, \cos \left(\frac{r-r_0}{\rho}\right) \right]\ ,
\eeq
which is valid for $|r-r_0|< \frac{\pi\,\rho}{2}$, and one can conveniently choose $r_0=0$. These solutions can be obtained from those of the preceding case letting
\beq
r - r_0 \ \to \ i\left( r - r_0 \ - \ \frac{\pi}{2}\right) \ , \qquad \Delta \ \to \ - \ i\, \Delta \ .
\eeq
 \item If $\epsilon=1$ and $E=0$, eq.~\eqref{conserv} reduces to
 \beq
 Z' \ = \ \pm \Delta\, e^{\,Z} \ ,
 \eeq
 which can be simply integrated and yields
 \beq
 e^{-Z} \ = \ \mp \, \Delta \left(r - r_0\right) \ ,
 \eeq
 where $r_0$ is another integration constant. Here clearly the upper sign is associated to the region $r< r_0$, where the real solution reads
 \beq
 Z \ = \ - \ \log \Delta \left( r_0 - r \right) \ .
 \eeq
 while the lower one is associated to the region $r > r_0$, where the real solution reads
 \beq
 Z \ = \ - \ \log \Delta \left( r - r_0\right) \ .
 \eeq
 These types of solutions capture the limiting behavior of the preceding cases when $E$ is negligible with the respect to the exponential $e^{2 Z}$, and thus as $\rho \to \infty$.
\item Finally, If $\epsilon = -1$ and $E= \frac{1}{\rho^2} > 0$, letting
\beq
 x \ = \ \frac{r}{\rho}
 \eeq
 eq.~\eqref{conserv} becomes
\beq
 \left(\frac{d Z}{d x}\right)^2  \ = \ 1 \ - \ \left(\Delta\,\rho\right)^2 \, e^{2\, Z} \ .
\eeq
The redefinition
\beq
Z = \ \tilde{Z} - \log(\Delta\, \rho)
\eeq
leads to the reduced equation
\beq
 \left(\frac{d \tilde{Z}}{d x}\right)^2  \ = \ 1 \ - \ e^{2\, \tilde{Z}} \ ,
\eeq
and therefore finally to the solution
\beq
Z \ = \ -  \log\left[\Delta\,\rho \, \cosh \left(\frac{r-r_0}{\rho}\right) \right]\ ,
\eeq
for all real values of $r$. This can be continued to the first solution combining the two transformations
\beq
\frac{r}{\rho} \ \to \ \frac{r}{\rho} \ + \ i \frac{\pi}{2} \ , \qquad \Delta \ \to \ - \ i\, \Delta \ .
\eeq
 \end{enumerate}
\end{appendices}

\end{document}